\documentstyle[12pt, epsf]{article}
\begin{document}

\begin{center}
{\large \bf 
Bosons in a Lattice: 
}
\end{center}
\begin{center}
{\large \bf
Exciton\,-\,Phonon Condensate in Cu$_{2}$O 
} 
\end{center}
\vspace*{0.3cm}
\begin{center}
 D. Roubtsov and Y. L\'epine
\end{center}
\begin{center}
{\it 
Groupe de Recherche en Physique et Technologie des Couches Minces, 
}
\end{center}
\begin{center}
{\it 
D\'epartement de Physique, Universit\'e de Montr\'eal, 
}
\end{center}
\begin{center}
C.P. 6128, succ. centre-ville, Montreal, PQ, H3C\,3J7, Canada 
\end{center}
\begin{center}
e-mail: roubtsod@physcn.umontreal.ca 
\end{center}
\vspace*{0.5cm}
\begin{abstract}

We explore a nonlinear field model to describe the interplay 
between  
the ability of  excitons to be  Bose-condensed  
and   
their interaction  with other modes of a crystal. 
We apply our consideration to the long-living 
para-excitons in  Cu$_{2}$O.
Taking into account the exciton-phonon interaction and introducing   
a coherent phonon part of the moving condensate, we derive the dynamic 
equations for the exciton-phonon condensate. 
These equations can support localized  solutions,
and  we discuss the conditions
for the moving inhomogeneous condensate to appear in the crystal.   
We calculate the condensate wave function and energy, and 
a collective excitation spectrum in the semiclassical approximation;
the inside-excitations were found to follow  
the asymptotic behavior of the macroscopic wave function exactly.  
The stability conditions of the moving condensate are analyzed by use of
Landau arguments, and Landau critical parameters appear in the theory.
Finally, we  apply our model to describe
the recently observed interference and strong nonlinear interaction between 
two coherent exciton\,-\,phonon 
packets in Cu$_{2}$O. 
\end{abstract}

\vspace*{0.5cm}



\newpage

\section{Introduction}

Excitons in semiconductor crystals \cite{review} and nanostructures 
\cite{Butov},\cite{Snoke} are a very interesting and challenging object to search
for the process of Bose-Einstein condensation (BEC).  
Nowadays there is a lot of experimental evidence that the optically 
inactive para-excitons in
Cu$_{2}$O can form a highly correlated state, or the excitonic 
Bose Einstein condensate 
\cite{review},\cite{Lin},\cite{Goto}.
A moving condensate of para-excitons in a
3D Cu$_{2}$O crystal turns out to be spatially  inhomogeneous  in the 
direction of
motion, and the registered velocities of coherent exciton packets 
turn out to be always less, but
approximately equal to the longitudinal sound speed of the crystal \cite{Fortin}.

Analyzing recent experimental \cite{Lin},\cite{Fortin},\cite{Benson}
and theoretical \cite{Hanamura1}-\cite{boser}
studies of BEC of excitons in Cu$_2$O, we can conclude that there are
essentially two different stages of this process. The first stage is
the kinetic one, with the characteristic time scale of $10\sim 20$\,ns.
 At this stage, a condensate of long-living para-excitons begins to be 
formed  from a quasi-equilibrium 
degenerate state of excitons ($\mu \ne 0$, $T_{\rm eff} > T_{\rm latt}$) 
when the concentration and the effective temperature 
of excitons in a cloud  meet the conditions of Bose-Einstein Condensation \cite{review}.   
Note that we do not discuss here the behavior of ortho-excitons (with the 
lifetime $\tau_{\rm ortho} \simeq 30$\,ns) and their influence on 
the para-exciton condensation process. For more details about the 
ortho-excitons
in Cu$_2$O, ortho-para-exciton conversion, etc.  see \cite{Lin},\cite{Goto},
\cite{Kavoulakis}, \cite{Lozovik}.  

The most intriguing feature of the kinetic stage is that formation of 
the para-exciton condensate 
and the process of momentum transfer to the para-exciton cloud are
happening simultaneously. If the diameter of an excitation spot 
on the crystal surface 
is large 
enough, \,$S_{\rm spot} \simeq S_{\rm surf}$, \,
and the energy of a laser beam  satisfies
$\epsilon_{\rm phot} \gg E_{\rm gap}$,    
nonequilibrium acoustic phonons
may  play 
the key role in 
the 
process of momentum transfer. As a result, the mode with macroscopical
occupancy of the excitons appears to be with   
$
\langle{\bf k}\rangle \ne 0, 
$
where 
$\hbar\langle k_{x} \rangle =  m_{\rm x}\, v$
and  $v$ is the packet velocity.

Indeed, the theoretical results obtained in 
the framework of the ``phonon wind'' model \cite{Tichodeev},\cite{discussion} 
and the experimental observations \cite{Lin},\cite{Goto},\cite{Fortin}
are the strong arguments in favor of this idea.
To the authors' knowledge, there are no realistic theoretical 
models of the kinetic stage of para-exciton condensate formation  where  
quantum degeneracy of the appearing exciton state and possible coherence of 
nonequilibrium phonons pushing the excitons would be taken into account.
Indeed, the condensate formation and many other processes involving it are essentially
nonliner ones. Therefore, the condensate, or, better, the 
{\it  macroscopically occupied mode}, can be different from $ n({\bf k}=0) \gg 1$, and the language of the states in
${\bf k}$-space and their occupation numbers $ n({\bf k})$ may be not relevant to 
the problem, see \cite{Ivanov}.
 
In this study, we will not explore the stage of condensate 
formation. Instead, we investigate 
the second, quasi-equilibrium stage, in which the condensate has already been 
formed and it moves through a crystal with some constant velocity and 
characteristic shape of the density profile.
In theory, the time scale of this ``transport'' stage,
$\Delta t_{\rm tr}$,
 could be 
determined by the para-exciton lifetime 
($\tau_{\rm para} \simeq 13\,\,\mu$s \cite{review}). 
In practice, it is determined by the characteristic size $\ell$ of a 
high-quality single crystal available for experiments: 
 $$
\Delta t_{\rm tr} \simeq \ell/c_{l}\simeq 0.5-2\,\mu{\rm s}   
\ll \tau_{\rm para}, 
$$
where $c_{l}$
is the longitudinal sound velocity. 
        
We assume that at the ``transport'' stage, 
the temperature of the moving packet 
(condensed\,$+$\,noncondensed particles) is approximately equal to the lattice 
 temperature, $$ 
T_{\rm eff}= T_{\rm latt}<T_{c}.$$
Then we can consider  the simplest case of $T=0$ and disregard the influence of all sorts
of {\it nonequilibrium} phonons (which appear at the stages of exciton 
formation, thermalization \cite{Tichodeev}
) on 
the formed moving condensate.

Any theory of the exciton BEC in Cu$_{2}$O has to point out some physical mechanism(s)
by means of which the key experimental facts can be explained. (For example,  
the condensate moves without friction
within a narrow interval of velocities localized near $c_{l}$, 
and   
the shape of the stable macroscopic wave function of excitons
resembles soliton profiles \cite{Benson}.) 
Here we explore a simple model of the ballistic exciton-phonon condensate.
In this case, the general structure of the Hamiltonian of the moving
exciton packet and the lattice phonons is the following:  
\begin{equation}
\hat{H}=H_{\rm ex}(\hat{\psi}^{\dag},\hat{\psi}) 
        - {\bf v}{\bf P}_{\rm ex}(\hat{\psi}^{\dag},\hat{\psi})
        + H_{\rm ph}(\hat{{\bf u}}, \hat
{\pi}
)
        - {\bf v}{\bf P}_{\rm ph}(\hat{{\bf u}}, \hat
{\pi}
 )
        + H_{\rm int}(\hat{\psi}^{\dag}\hat{\psi},\,\partial_{j}\hat{ u}_{k}). 
 \label{ham1}
\end{equation}
 Here $\hat{\psi}$ is the Bose-field operator describing the excitons,
$\hat{\bf u}$ is the field operator of lattice displacements,  
$\hat
{\pi}
$ 
is the momentum density operator canonically 
conjugate to $\hat{\bf u}$, 
and ${\bf P}$ is  the momentum operator. Note 
that the Hamiltonian (\ref{ham1}) is 
written in the reference frame moving with the exciton packet, i.e.
$ {\bf x} \rightarrow {\bf x} - {\bf v}t$ and ${\bf v} = {\rm 
const}$ is the ballistic velocity of the packet.     

\section{3D Model 
of Moving Exciton-Phonon Condensate}

To derive the equations of motion of the field operators (and generalize these equations to the case of  
$T \ne 0 $), it is more convenient to start from the Lagrangian.
In the proposed model, the Lagrangian density has the following form
in the co-moving frame
$$
{\cal L}={{i\hbar}\over{2}}( \hat{\psi}^{\dag}\,\partial_{t}\hat{\psi} - 
                    \partial_{t}\hat{\psi}^{\dag}\,\hat{\psi}) 
 +\,v{{i\hbar}\over{2}}(\partial_{x}\hat{
\psi}^{\dag}\,\hat{\psi} - 
\hat{\psi}^{\dag}\partial_{x}\hat{\psi}) -
$$
$$
- \tilde{E}_{g}\hat{\psi}^{\dag}\hat{\psi} - {{\hbar^{2}}\over{2m}}
\nabla
\hat{\psi}^{\dag}\nabla\hat{\psi}
- {{\nu_{0}}\over{2}} \bigl(\psi^{\dag}({\bf x},t)\bigr)^{2}
\bigl(\psi({\bf x},t)\bigr)^{2} - {{\nu_{1}}\over{3}}\bigl(\psi^{\dag}({\bf 
x},t)\bigr)^{3}\bigl(\psi({\bf x},t)\bigr)^{3} +
$$
$$
+ {{\rho}\over{2}}(\partial_{t}\hat{\bf u})^{2} -{{\rho c_{l}^{2}}\over{2}} 
(\,\partial_{j}\hat{u}_{s}\,)^{2}
-{{\rho c_{l}^{2}}\over{2}}\,\kappa_{3}\, 
(\partial_{j}\hat{u}_{s}\,)^{3}
-\frac{ \rho v}{2}\,(\partial_{t}\hat{\bf u}\,\partial_{x}\hat{\bf u}  +  \partial_{x}\hat{\bf u}\,  \partial_{t}\hat{\bf u}\,)     
+
{{\rho v^{2}}\over{2}}(\partial_{x}\hat{\bf u})^{2}-
$$
\begin{equation}
-\sigma_{0} \hat{\psi}^{\dag}({\bf x},t)\hat{\psi}({\bf x},t)
\nabla\hat{\bf u}({\bf x},t),
\label{lagrangian}
\end{equation}
where  $m$ is the exciton ``bare''mass ($m= m_{e}+m_{h}\simeq 3m_{e}$
for $1$s excitons in Cu$_{2}$O),  \,$\nu_{0}$ is the exciton-exciton 
interaction
constant ($\nu_{0}>0$ corresponds to the repulsive interaction between 
para-excitons in Cu$_{2}$O \cite{Schmitt-Rink}), $\rho$ is the crystal density,  
$\sigma_{0}$ is the exciton-longitudinal phonon coupling constant, 
and ${\bf v}=(v,0,0)$.  
The energy of a free exciton is $\tilde{E}_{g} + \hbar^{2}{\bf k}^{2}/2m$.   
Although the validity of the condition 
$n\,\tilde{a}_{\rm B}^{3} \ll 1$ 
($\tilde{a}_{\rm B}$ is the exciton Bohr radius)
makes it possible to disregard
all the multiple-particle interactions with more than two participating 
particles in $\hat{H}_{\rm ex}$ \cite{Keldysh}, we include the hard-core repulsion term
originated from 
the three-particle interaction in $\cal L$, i.e. 
$\nu_{1} \ne 0$\, 
and \,
$$
0< \nu_{1}/\tilde{a}_{\rm B}^{6} \ll \nu_{0}/\tilde{a}_{\rm B}^{3} \simeq  {\rm const}\,{\rm Ry}^{*}.
$$
(For 3D case, one has to take 
\,${\rm const} \simeq 10$ because \,$ \nu_{0} = 4\pi\,(\hbar^{2}/m)\,a_{\rm sc}$ 
and $a_{\rm sc} \simeq (1\sim 3)\,\tilde{a}_{\rm B}$; 
see the discussion in \cite{ScLength}.)

Moreover, in the Lagrangian of the displacement field,  
we include the first nonlinear term  
\,$ \propto \kappa_{3}\,(\partial u)^{3}$. 
(The  dimensionless parameter $\kappa_{3}$ 
originates from Taylor's expansion   
of an interparticle potential \,$U(\vert r_{i} - r_{j} \vert )$ \,
of the medium atoms.)  
Assuming that a dilute excitonic packet 
moves in a weekly nonlinear medium,  we will not take into 
account  more higher nonlinear terms in  (\ref{lagrangian}). 

For simplicity's sake, we 
take
all the interaction terms in ${\cal L}$ in the {\it local} form and 
disregard the interaction between the excitons and transverse phonons
of the crystal.     
Note that the  ballistic  velocity $v$
is one of the parameters of the theory, and we will not 
take into account  
the excitonic normal component and velocity, i.e. 
$ v = v_{s} \sim \nabla \varphi_{\rm c}$,
\, $(T=0)$.   
This means that we choose the spatial part of the coherent phase of the packet, 
$\varphi_{\rm c}({\rm x})$, to be in the simplest form,  
\begin{equation}
\exp\bigl(i\varphi_{\rm c}({\rm x})\bigr)=\exp\bigl(i(\varphi + k_{0}x)\bigr), 
\,\,\,\,\,\varphi={\rm const},\,\,\,\,\,\hbar k_{0}=mv. 
\label{superF}
\end{equation}

The equations of motion can be easily derived by the standard variational
method from the following condition: 
$$
\delta S = \delta \!\int dt\,d{\bf x}\,
{\cal L}\bigl(\hat{\psi}^{\dag}({\bf x},t),\,\hat{\psi}({\bf x},t), \,\hat{\bf u}({\bf x},t)
\bigr)=0.  
$$
Indeed, after transforming the Bose-fields $\hat{\psi}^{\dag}$ and  
$\hat{\psi}$ by 
$$
\hat{\psi}({\bf x},t) \rightarrow \exp(-i\tilde{E}_{g}t/\hbar)
\exp (imvx/\hbar)\hat{\psi}({\bf x},t),
$$
we can write these equations as follows:
$$
(i\hbar\partial_{t} + mv^{2}/2) \hat{\psi}({\bf x},t)= 
$$
\begin{equation}
=\Bigr(- {{\hbar^{2}}\over{2m}}\Delta + \nu_{0}
\hat{\psi}^{\dag}\hat{\psi}({\bf x},t)+ 
\nu_{1}\hat{\psi}^{\dag \,2}\hat{\psi}^{2}({\bf x},t)\,\Bigl)  
\hat{\psi}({\bf x},t) + \sigma_{0}\nabla\hat{\bf u}({\bf x},t)\,
\hat{\psi}({\bf x},t),
\label{eq1}
\end{equation}
\begin{equation}
\bigl(\partial_{t}^{2}- c_{l}^{2}\Delta - v(\partial_{t}\partial_{x} + \partial_{x}\partial_{t})
+ v^{2}\partial_{x}^{2}\bigr)\hat{u}_{s}({\bf x},t)
-c_{l}^{2}  \sum_{j}\!
2 \kappa_{3}\,\partial_{j}^{2} \hat{ u}_{s}\,\partial_{j}\hat{ u}_{s}({\bf x},t) 
=
\rho^{-1}\sigma_{0}\partial_{s}\bigl(\hat{\psi}^{\dag}\hat{\psi}({\bf x},t)\bigr).
\label{eq2}
\end{equation}

We  assume that the condensate of excitons {\it exists}. This  means that 
the following representation of the exciton Bose-field holds: 
 $\hat{\psi}=\psi_{0} + \delta\hat{\psi}$.
Here $\psi_{0} \ne 0$ is the classical part of the field operator $\hat{\psi}$
or, in other words, the condensate wave function, and  $\delta\hat{\psi}$
is the fluctuational part of $\hat{\psi}$, which describes 
out-of-condensate particles.

One of the important objects in the theory of BEC is the correlation functions
of Bose-fields. The standard way to calculate them in this model (the 
excitonic function $\langle \psi({\bf x},t)\psi^{\dag}
({\bf x}',t') \rangle $, for example,) can be based on the effective 
action or the effective 
Hamiltonian approaches \cite{Popov}.   Indeed, one can, first, integrate over the
phonon variables ${\bf u}$, get the expression for 
$ S_{\rm eff}(\psi, \psi^{\dag})$ and, second, use $S_{\rm eff} $ (or 
$\hat{H}_{\rm eff}$) to derive the equations of
motion for $\psi_{0}$, $\delta\hat{\psi}$, correlation functions, etc..   

In this work  we do not follow  that way; instead, we treat excitons and 
phonons equally \cite{Loutsenko},\cite{miscellaneous},\cite{Davydov}.
This means that 
the displacement field $\hat{\bf u}$ can have 
 a nontrivial coherent part too, i.e. $\hat{\bf u}={\bf u}_{0} + \delta\hat{\bf u}$ and
${\bf u}_{0} \ne 0$, and the actual moving condensate can be an exciton-phonon one,
i.e. $\psi_{0}({\bf x},t)\cdot {\bf u}_{0}({\bf x},t)$.  
Then the equation of motion for the classical parts of the fields $\hat{\psi}$ and $\hat{\bf u}$
can be derived by use of the variational method from 
${\cal L} ={\cal L}(\psi, \psi^{*}, {\bf u})$, in which all the fields can be considered as the classical ones.
Eventually we have 
$$
(\,i\hbar\partial_{t} + mv^{2}/2\,){\psi}_{0}({\bf x},t)= 
$$
\begin{equation}
=\Bigr(- {{\hbar^{2}}\over{2m}}\Delta + \nu_{0}
\vert{\psi_{0}}\vert^{2}({\bf x},t)+ 
\nu_{1}\vert{\psi}_{0}\vert^{4}({\bf x},t)\,\Bigl)  
{\psi}_{0}({\bf x},t) + \sigma_{0}\nabla{\bf u}_{0}({\bf x},t)\,
{\psi}_{0}({\bf x},t)\,,
\label{eq11}
\end{equation}
\begin{equation}
\bigl(\partial_{t}^{2}- c_{l}^{2}\Delta - 2v\partial_{t}\partial_{x} 
+ v^{2}\partial_{x}^{2}\bigr){u}_{0\,s}({\bf x},t)
-c_{l}^{2}  \sum_{j}\!
2 \kappa_{3}\,\partial_{j}^{2} { u}_{0\,s}\,\partial_{j}{ u}_{0\,s}({\bf x},t)
=
\rho^{-1}\sigma_{0}\partial_{s}\bigl(\vert\psi_{0}\vert^{2}({\bf x},t)\bigr).
\label{eq22}
\end{equation}
Notice that deriving these equations we disregarded the interaction between
the classical (condensate) and the fluctuational 
(noncondensate) parts of the fields. That is certainly a good approximation
for $T=0$ and $T \ll T_{c}$ cases \cite{Griffin}.

In this article, a steady-state of the condensate is the object of the main 
interest. In the co-moving frame of reference, 
the condensate steady-state is just the  stationary solution of Eqs. 
(\ref{eq11}),\,(\ref{eq22}) and it can be taken in the form
$$
\psi_{0}({\bf x},t)=\exp(-i\mu t)\exp(i\varphi)\phi_{\rm o}({\bf x}), 
\,\,\,\,\,
u_{0}({\bf x},t)= {\bf q}_{\rm o}({\bf x}),
$$ 
where  $\phi_{\rm o}$ and  ${\bf q}_{\rm o}$
are the real number functions, and $\varphi ={\rm const}$ is the coherent phase
of the condensate wave function in the co-moving frame, see Eq. (\ref{superF}).
(This phase can be taken equal zero if only a single condensate 
is the subject of interest.) 

Then,  the following equations have to be solved, 
(\,$\mu= \tilde{\mu} - mv^{2}/2 $\,):
\begin{equation}
\tilde{\mu}{\phi}_{\rm o}({\bf x})= 
\Bigr(- {{\hbar^{2}}\over{2m}}\Delta + \nu_{0}
\phi_{\rm o}^{2}({\bf x})+ 
\nu_{1}\phi_{\rm o}^{4}({\bf x})\,\Bigl)  
\phi_{\rm o}({\bf x}) + 
\sigma_{0}\nabla{\bf q}_{\rm o}({\bf x})\,
{\phi}_{\rm o}({\bf x})\,,
\label{eq111}
\end{equation}
\begin{equation}
- \bigl\{\,(c_{l}^{2}- v^{2})\partial_{x}^{2}
+ c_{l}^{2}\partial_{y}^{2} + c_{l}^{2}\partial_{z}^{2}\,\bigr\}
{ q}_{{\rm o}\,s}({\bf x})
-\,c_{l}^{2} \! \sum_{j=x,y,z}\!
2 \kappa_{3}\,\partial_{j}^{2} 
{ q}_{{\rm o}\,s}\,\partial_{j}{ q}_{{\rm o}\,s}({\bf x})
=
\rho^{-1}\sigma_{0}\partial_{s}\phi_{\rm o}^{2}({\bf x}).
\label{eq222}
\end{equation}
Note that in order to simplify Eq. (\ref{eq22}) to Eq. (\ref{eq222}), 
we assumed only
\,$u_{0}({\bf x},t)= {\bf q}_{\rm o}({\bf x})$. 
In this model, it is enough to obtain localized solutions for the
displacement field.

\section{Effective 1D Model for the Condensate Wave Function}

Solving Eqs. (\ref{eq111}),(\ref{eq222})
in 3D space  seems to be a difficult problem. 
However, these equations can be essentially simplified  if we 
assume that the condensate is inhomogeneous along the 
$x$-axis only, that is  
$$
\phi_{\rm o}({\bf x})=\phi_{\rm o}(x) 
\,\,\,\,\,{\rm and}\,\,\,\,\, 
{\bf q}_{\rm o}({\bf x})=(q_{\rm o}(x), \,0,0).
$$
Note that the cross-section area of an excitation spot, $S$, 
has  to be basically constant across the sample cross-section. 
In this case,  the problem 
can be considered as an effectively one-dimensional one.

Such an effective reduction of dimensionality 
transforms difficult  
(nonlocal differential) equations for the condensate wave function  
into a rather simple  
differential ones, and obtained in this way the    
 effective 1D model for the condensate wave function  
$\phi_{\rm o}\cdot  q_{\rm o}$ \,conserves 
all the important properties       
of the ``parent'' 3D model.

Indeed, if $v<c_{l}$, the following equations 
stand for the condensate ($y(x)=\partial_{x}q_{\rm o}(x)$):
\begin{equation} 
\tilde{\mu}\phi_{\rm o}(x)=\bigl(\, -(\hbar^{2}/2m) 
\partial_{x}^{2} +\nu_{0}\phi_{\rm o}^{2}(x) +
\nu_{1}\phi_{\rm o}^{4}(x)\,\bigr)\phi_{\rm o}(x) + \sigma_{0}\,y(x)\,\phi_{\rm o}(x),
\label{1Deq} 
\end{equation}
\begin{equation}
-(c_{l}^{2} - v^{2})\partial_{x} y(x)  - 2 c_{l}^{2}\kappa_{3}\,\partial_{x} y\,y(x) = \rho^{-1}\sigma_{0}
\,\partial_{x} \phi_{\rm o}^{2}(x).
\label{11Deq}
\end{equation}
The last equation can be easily integrated,
\begin{equation}
y(x) + \tilde{\kappa}_{3}\, y^{2}(x)  =  \Phi(x) + {\rm const}, 
\label{11DeqD}
\end{equation}
and solved relative to $y(x)$. Here 
$$
\tilde{\kappa}_{3}=\frac{c_{l}^{2}}{c_{l}^{2}-v^{2}}\,\kappa_{3}
\equiv \gamma(v)\,\kappa_{3},\,\,\,\,\,\,\,
\Phi(x)= - \frac{ \sigma_{0} }{ \rho (c_{l}^{2}-v^{2}) }
\,\phi_{\rm o}^{2}(x) \equiv -\gamma(v)\,
\frac{ \sigma_{0} }{ \rho\,c_{l}^{2} }\,\phi_{\rm o}^{2}(x).
$$
Note that  the  medium nonlinearity parameter
 $\kappa_{3}$ 
\,can be enhanced by the factor of the order of $4 \sim 10$
if the value of \,$v$\, is less, but close to \,$c_{l}$.
(For spatially localized solutions, \,$\partial_{x}q_{\rm o}(x) \simeq 0$
and $\phi_{\rm o}^{2}(x) \simeq 0$ at $\vert x \vert \gg L_{\rm ch}$, 
so that \,${\rm const}=0$.)  

If $\kappa_{3}<0$,
we can always represent  the solution of (\ref{11Deq}),(\ref{11DeqD})
in the following form
\begin{equation}
y(x)=  \Phi(x) + \vert \tilde{\kappa}_{3} \vert \,\Phi^{2}(x) + 
2 \, \tilde{\kappa}_{3}^{2}\,\Phi^{3}(x)+\cdots.
\label{dc}
\end{equation}
(Indeed,  the parameters of  medium nonlinearity 
can be chosen as 
\,$\kappa_{3}<0$\, and \,$\kappa_{4}>0$
\cite{Hu}.)
After substitution of (\ref{dc}) into Eq. (\ref{1Deq}), 
Eqs. (\ref{1Deq}), (\ref{11Deq})  can be rewritten  as follows:
\begin{equation}
\tilde{\mu}\,\phi_{\rm o}(x)=\bigl(\, -(\hbar^{2}/2m)\,
\partial_{x}^{2} +\widetilde{\nu_{0}}\,\phi_{\rm o}^{2}(x)+
\widetilde{\nu_{1}}\,\phi_{\rm o}^{4}(x) + \epsilon_{2}\,\bigr)\phi_{\rm o}(x),
 \label{1DeqMod} 
\end{equation}
\begin{equation}
\partial_{x}\,q_{\rm o}(x)  =  \Phi(x) + \vert \tilde{\kappa}_{3} \vert \,\Phi^{2}(x) + 
\epsilon_{2}',
\label{dcNL}
\end{equation}
where  the interparticle interaction constants are renormalized as follows
\begin{equation}
\widetilde{\nu_{0}} = \nu_{0} \,-\,\sigma_{0}\,\frac{ \sigma_{0}}{ \rho (c_{l}^{2}-v^{2}) }, \,\,\,\,\,\,
\widetilde{\nu_{1}} =
\nu_{1}\, + \,\sigma_{0}\,
\frac{ \sigma_{0}^{2}}{ \bigl(\,\rho (c_{l}^{2}-v^{2})\,\bigr)^{2} }\,
\vert \tilde{\kappa}_{3} \vert, 
\label{kappa} 
\end{equation}
and  more higher nonlinear terms  are designated  by  $\epsilon_{2}$.
A small parameter in  Eq. (\ref{kappa}) comes from
the term 
$$
\sigma_{0} / \rho(c_{l}^{2}-v^{2}) =
 \gamma(v)\,\bigl(\,\sigma_{0} / M c_{l}^{2}\,\bigr)\,a_{l}^{3},
$$ 
where \,$\gamma(v)=  c_{l}^{2}/(c_{l}^{2}-v^{2})$ and 
$M$ is the mass of the crystal elementary cell.

The effective two-particle interaction constant $\widetilde{ \nu_{0} }\,(v)$ \,can be {\bf negative}
if the velocity of the condensate lies inside the interval\, $v_{\rm o}<v<c_{l}$, 
 where 
\begin{equation}
v_{\rm o}=\sqrt{c_{l}^{2}-(\sigma_{0}^{2}/\rho\nu_{0}) }.
\label{velo1}
\end{equation}
Outside this interval, \,$\widetilde{\nu_{0}}\,(v) > 0$ \cite{Loutsenko}
and the velocity  $v_{\rm o}$  can be called  the first
`critical' velocity in the model.
The meaning of this   velocity  can be clarified by 
rewriting (\ref{kappa}) in the dimensionless form, 
\begin{equation}
\frac{ \widetilde{ \nu_{0} }}{ \sigma_{0}\,a^{3}_{l}} \,= \,\frac{ \nu_{0}}{ \sigma_{0}\,a^{3}_{l}}
-
 \gamma(v) \left( \frac{ \sigma_{0}}{ M c_{l}^{2}  } \right),
\end{equation}
If $v > v_{\rm o}$, 
\begin{equation}
\gamma(v) \left( \frac{ \sigma_{0}}{ M c_{l}^{2}  } \right)\,
>\, \frac{ \nu_{0}}{ \sigma_{0}\,a^{3}_{l}}\,\simeq \,\frac{{\rm const}\, {\rm Ry}^{*}}{ \sigma_{0}}\,
\frac{ \tilde{a}_{B}^{3}}{ a^{3}_{l} },
\label{negative}
\end{equation}
where  ${\rm Ry}^{*}$\, and \,$\tilde{a}_{B}^{2} $ 
are the characteristic energy and the cross-section
of  two-particle collisions  in the exciton subsystem.
The following inequalities are true  for excitons in a crystal 
$$
\tilde{a}_{B}^{3} > (\gg)\, a^{3}_{l}\,\,\,\,\,{\rm and} \,\,\,\,\,\
{\rm const \, Ry}^{*} < (\ll)\, \sigma_{0}, 
$$ 
and, usually,  the value of the parameter $\nu_{0}/ \sigma_{0}\,a^{3}_{l} $ 
\,  is \,$> 1$.  

For para-excitons in Cu$_{2}$0, however, we assume the (effective) value
of \,$ \nu_{0} / \sigma_{0}\,a^{3}_{l}$ 
can be estimated as \,$0.3 \sim 0.6\, <\,1$,   \,whereas   
the value of  $\sigma_{0} / M c_{l}^{2} \simeq  0.1 \sim  0.3$.
This makes the inequality 
(\ref{negative}) 
valid at, say,  $ v \approx (0.8 \sim 0.9)\, c_{l}$, or \,$\gamma(v) \simeq 5$.
Thus, within the effective 1D model,
the critical factor $\gamma_{\rm o}= \gamma( v_{\rm o})$ 
is the following ratio
$$  
\gamma_{\rm o}= \left( \frac{ \nu_{0}}{ \sigma_{0}\,a^{3}_{l}}\right ) / \left(
\frac{ \sigma_{0}}{ M c_{l}^{2}} \right ),
$$
and, for the substances with  $\nu_{0} / \sigma_{0}\,a^{3}_{l} <1$,  the regime with $\widetilde{ \nu_{0} } < 0$
can be obtained at velocities {\it reasonably close} 
but not equal to $c_{l}$, for example,  beginning from  some  
\,$ \gamma_{\rm o} < 10$, \,
(\,$ \gamma (\,0.95\,c_{l}) \approx 10$\,). 

On the other hand,  
the effective three-particle interaction constant 
\,$\widetilde{ \nu_{1}}\,(v)$\,
is always  positive for  crystals with $\kappa_{3}< 0$. It can be represented in the dimensionless form
as follows  
\begin{equation}
\frac{ \widetilde{ \nu_{1} } }{ \sigma_{0} (a^{3}_{l})^{2} } 
\simeq 
{\rm const}'\, \left(\frac{ \nu_{0}}{ \sigma_{0}\, a^{3}_{l}} \right)\,\frac{\tilde{a}_{B}^{3}}{a^{3}_{l}}\,+\,
\gamma(v)\,\vert \kappa_{3} \vert 
\left( \gamma(v)\,\frac{ \sigma_{0}}{ M c_{l}^{2}  } \right)^{2}.
\label{three-part}
\end{equation}
Here we  estimated the 
``bare'' vertex of the three-particle collisions as   
$$
\nu_{1} \simeq  {\rm const}\, {\rm Ry}^{*} \tilde{a}_{B}^{6} 
\simeq  {\rm const}'\, \nu_{0}\, \tilde{a}_{B}^{3},  
\,\,\,\,\,\,{\rm const}' \le 1,
$$ 
and the same ${\rm Ry}^{*} $ can be  taken 
as a characteristic energy of collisions. 
The effective vertex \,$\widetilde{ \nu_{1} }>0$\, is enhanced by the
medium nonlinearity, and the both terms in the r.h.s. of 
(\ref{three-part})
can be equally important at $\gamma(v)>\gamma_{\rm o} $. 

Note that in the case of {\it strongly nonlinear lattices} with excitons, 
the effective interaction vertices in  (\ref{1DeqMod})     
(\,$\widetilde{ \nu_{1} }$,\,  $\widetilde{ \nu_{2} }$, \,etc.) 
depend on the velocity $v$ and the parameters of medium nonlinearity
 ($\kappa_{3}$, \,$\kappa_{4}$, etc.).
Then the effective exciton-exciton interaction   
can be strongly renormalized at 
sufficiently large gamma-factors
$\gamma(v)$ and
the vertices may change their signs
as it can happen with \,$\widetilde{ \nu_{0} }\,(v)$. 
\,In this article, however, we consider the case of    
{\it weakly nonlinear medium} with excitons (e.g.,   
a crystal with long living excitons).
More accurately, this means that at velocities $v \rightarrow c_{l} $
the effective vertex $\widetilde{ \nu_{0} }\,(v)$ became $<0$, while
the more higher vertices, such as 
$\widetilde{ \nu_{1} }\,(v)$ and   $\widetilde{ \nu_{2} }\,(v)$,
do not change their sign;  they remain   
\,$>0$ at \,$\gamma(v) > \gamma_{0}$.
Finally, to describe the weakly nonlinear case, it is enough to
take into account the parameters \,$\nu_{1}>0$\, 
and \,$\kappa_{3}<0$  and neglect more higher nonlinearities ($\epsilon_{2}$ and $\epsilon_{2}'$ in 
(\ref{1DeqMod}), (\ref{dcNL})\,).

In this study, we will consider the case of $v_{\rm o}<v<c_{l}$ in 
detail.
Indeed, in the case of  \,$\widetilde{ \nu_{0} }\,(v)< 0$ 
and  \,$\widetilde{ \nu_{1} }\,(v) > 0$, some localized solutions  
of Eqs. (\ref{1DeqMod}),(\ref{dcNL}) do exist. 
For example, the so-called  `bright soliton' solution of (\ref{1DeqMod})
exists if the generalized chemical potential is negative,  
\,$\tilde{\mu} <0$,  and  \,$\vert \tilde{\mu} \vert < \mu^{*}$.
Here 
\begin{equation}
 \mu^{*}=\frac{ \vert \widetilde{ \nu_{0} }  \vert^{2} }{(16/3)\,
\widetilde{ \nu_{1} } } \approx 
0.2\,\sigma_{0}\, \frac{\bigl(\,\vert \widetilde{ \nu_{0} }  \vert /
\sigma_{0}\,a_{l}^{3}\,\bigr)^{2} }
{\bigl(\,\widetilde{ \nu_{1} } /\sigma_{0}\,a_{l}^{6}\,\bigr) }.
\end{equation}
For \,$ \vert\kappa_{3} \vert  \sim 1 $ and \,$\gamma(v) > \gamma_{\rm o}\simeq  3 \sim 5$, 
\,we can roughly estimate the effective vertex  $\widetilde{ \nu_{1} }\,(v)$ as 
$$
\widetilde{ \nu_{1} }\,(v)/ \sigma_{0}\,a_{l}^{6} \simeq  (1 \sim 10)\,\bigl(\nu_{0}/\sigma_{0}\,a_{l}^{3}\bigr).
$$
Then \,$\mu^{*}(v) \simeq (10^{-1} \sim 10^{-2})\,{\rm Ry}^{*}$, 
and the more is the value of $\vert \kappa_{3} \vert $ the less is the value of $\mu^{*}(v)$. 

The `bright soliton' solution of Eq. (\ref{1DeqMod}) can be represented in the following form
\begin{equation}
\phi_{\rm o}(x)= \Phi_{\rm o} \,f\bigl(\,\beta(\Phi_{\rm o})\, x, \,\eta_{1}(\Phi_{\rm o})\,\bigr),\,\,\,\,\,\,
\beta(\Phi_{\rm o}) =\sqrt{ \frac{2 m}{ \hbar^{2} }\,\vert \tilde{\mu} \vert(\Phi_{\rm o}) }. 
\label{GENform}
\end{equation}
Here $\eta_{1}(\Phi_{\rm o})$ is some dimensionless parameter,  
and the generalized chemical potential \,$\tilde{\mu}<0$ is given by the formula  
\begin{equation} 
\vert \tilde{\mu} \vert =\vert \tilde{\mu} \vert(\Phi_{\rm o}) =
\vert \widetilde{\nu_{0}} \vert \, \Phi_{\rm o}^{2}/ 2  - 
\widetilde{\nu_{1}}  \, \Phi_{\rm o}^{4}/ 3 .
\label{muNEW}
\end{equation}
Like the chemical potential $\vert \tilde{\mu} \vert$, 
the amplitude of the bright soliton, $\Phi_{\rm o}$,  satisfies 
$$
\Phi_{\rm o}^{2} < (\Phi_{\rm o}^{*})^{2}= 
\vert \widetilde{ \nu_{0} } \vert /(\,4/3\,\widetilde{ \nu_{1} }),
$$
and 
\,$ \mu^{*} = \vert \widetilde{ \nu_{0} }  \vert \,\Phi_{\rm o}^{*\,2}/4$.

For $\vert \tilde{\mu} \vert /\mu^{*} \ll 1 $, the following approximation is valid 
\begin{equation}
\eta_{1} \approx \frac{1}{4}\,\bigl(\vert \tilde{\mu} \vert / \mu^{*}\bigr) + 
\frac{1}{8}\,\bigl(\vert \tilde{\mu} \vert / \mu^{*}\bigr)^{2} \ll 1,
\label{etaAPPR}
\end{equation}
and this  formula  can be used  up to
\,$\vert \tilde{\mu} \vert /\mu^{*} \simeq 0.5 $.
Then we can {\it approximate} the solution of  (\ref{1DeqMod}) by the following formulas
\begin{equation}
\phi_{\rm o}(x) \approx \Phi_{\rm o}\,\Bigl(\sqrt{1-\eta_{1}(\Phi_{\rm o})}\,
\cosh(\beta(\Phi_{\rm o})\,x)\,+\,\bigr(1-\sqrt{1-\eta_{1}(\Phi_{\rm o}) }\bigl)\,\Bigr)^{-1},     
\label{SolitonMod}
\end{equation}
\begin{equation}
\phi_{\rm o}(x) \simeq 2 \Phi_{\rm o}\, {\rm exp}(-\beta(\Phi_{\rm o})\,\vert x\vert\, ) / \sqrt{ 1-\eta_{1}}
\,\,\,\,\,\,\,{\rm for}\,\,\,\,\,\,\, \vert x \vert >  2 \beta(\Phi_{\rm o})^{-1},
\,\,\,\,\,\vert \mu \vert \ll \mu^{*}.
\end{equation}
The amplitudes of the exciton and phonon parts of the condensate, 
the characteristic width of the 
condensate, 
and the value of the 
effective chemical 
potential $\tilde{\mu}$ depend on the normalization of the exciton wave 
function $\phi_{\rm o}(x)$. We normalize it in 3D space assuming that the 
characteristic width of the packet in the $(y,z)$-plane is sufficiently large, i.e.
the cross-section area of the packet $S_{\perp}$  
can be made equal to  
the cross-section area S of a laser beam 
and  
$$
S_{\perp} \simeq S \simeq S_{\rm surf}.
$$
Then we can write this condition as follows:   
\begin{equation}
\int\!\!\vert\psi_{0}\vert^{2}(x,t)\,d{\bf x}
=S\int\!\!\phi_{\rm o}^{2}(x)\,dx = N_{\rm o},
\label{norma}   
\end{equation} 
where $N_{\rm o}$ is the number of condensed excitons, and, generally, 
$N_{\rm o} \ne N_{\rm tot}$.


Applying this  normalization condition, 
we get the following results 
\begin{equation}
\Phi_{\rm o}^{2} \approx 
\frac{ \vert \widetilde{\nu_{0}}\,(v)\vert }{  
2 \left( N_{\rm o}^{*} / N_{\rm o}  \right )^{2} x\,{\rm Ry}^{*}\,\tilde{a}_{B}^{6}
\,+\, 2\,\widetilde{\nu_{1}}\,(v)   }.
\label{AMP1} 
\end{equation}
Here we used the following notations,
$N_{\rm o}^{*}= 2 S /\tilde{a}_{B}^{2}$, 
$\hbar^{2}/m=2 x\,{\rm Ry}^{*}\, \tilde{a}_{B}^{2}$, where 
${\rm Ry}^{*}=\hbar^{2}/2\mu_{\rm exc}\tilde{a}_{B}^{2}$ 
\,and \,$x=\mu_{\rm exc}/m$.
The formula (\ref{AMP1}) is valid for \,$\vert \tilde{\mu}\vert/ \mu^{*} < 0.3\sim 0.4$.
We assume that,  at \,$N_{\rm o}^{*}/N_{\rm o} = \bar{n}_{\rm o} > 10$
(this is the important parameter!), we always have
$$
2\,\bar{n}_{\rm o}^{2}\,x\,{\rm Ry}^{*}\,\tilde{a}_{B}^{6} \,\gg\, 
\widetilde{\nu_{1}}\,(v) = \tilde{\varepsilon}_{1}( \vert \kappa_{3}\vert, v) \,\tilde{a}_{B}^{6} \simeq 
(1\sim 10)\,{\rm Ry}^{*}\,\tilde{a}_{B}^{6}.
$$
Then, the following inequalities are valid: 
$\Phi_{\rm o}^{2}(N_{\rm o},\,v) \ll \Phi_{\rm o}^{*\,2}$ \,and 
\begin{equation}
\vert \tilde{\mu} \vert (N_{\rm o},\,v) \approx 
\frac{ \vert \widetilde{\nu_{0}}\,(v)\vert ^{2}}{ 2 \left\{
2\,\bar{n}_{\rm o}^{2}\,x\,{\rm Ry}^{*}\,\tilde{a}_{B}^{6} \,+\,
4 \,\widetilde{\nu_{1}}\,(v) \, \right\}   } \ll \mu^{*}= 
 \frac{ \vert \widetilde{\nu_{0}}\,(v)\vert^{2} }{ 5.3\, \widetilde{\nu_{1}}\,(v)    }.
\label{MU1}
\end{equation}

The characteristic length of the packet can be estimated from Eq. (\ref{GENform}) as follows
($\widetilde{\nu_{0}}\,(v) = \tilde{\varepsilon}_{0}(v) \,\tilde{a}_{B}^{3}$):
$$
L_{\rm ch}^{-1}(N_{\rm o},\,v) \simeq \frac{1}{ 4}\, \beta  (\Phi_{\rm o}) \approx
\frac{1}{ 4} \frac{ \vert \tilde{\varepsilon}_{0}(v)\vert    }{( 
2\,x\,{\rm Ry}^{*}\, \tilde{a}_{B}) }   \, \frac{1}{\sqrt{ \bar{n}_{\rm o}^{2} \,+\, 
\bigl(\tilde{\varepsilon}_{1}(v) /x\,{\rm Ry}^{*} \bigr)   }   }\,\simeq
$$
\begin{equation}
\simeq \, \frac{1}{ 8}\,
\frac{ \vert \tilde{\varepsilon}_{0}(v)\vert   }{x\,{\rm Ry}^{*}}\, \frac{1}{\tilde{a}_{B}\, \bar{n}_{\rm o}
  } \,\,\,\,{\rm at}\,\,\,\,\bar{n}_{\rm o}>10.
\label{Width}
\end{equation}
Therefore, at \,$\gamma(v) \simeq 2\gamma_{\rm o}$, we can roughly estimate 
\begin{equation}
L_{\rm ch}(N_{\rm o},\,v) 
\simeq  4\,\frac{{\rm Ry}^{*} }{ \vert \tilde{\varepsilon}_{0}(v)\vert }
\,\bar{n}_{\rm o}\,\tilde{a}_{B}  \sim   4\,(10^{1}\,\bar{n}_{\rm o})\,\tilde{a}_{B}, 
\end{equation}
and, 
for the average concentration of condensed excitons in the packet, $n_{\rm o}$,
we have  
$$ 
n_{\rm o}\,
\tilde{a}_{B}^{3} \approx (N_{\rm o}/SL_{\rm ch})\,\tilde{a}_{B}^{3} \simeq 1 /\bar{n}_{\rm o}^{2} \ll 1. 
$$ 

Recall that the second part of the condensate, 
the displacement field $q_{\rm o}(x)$, is of the same importance as the first part, 
the  exciton wave function $\phi_{\rm o}(x)$.
The displacement field  $\partial_{x} q_{\rm o}(x)$\,
can be represented as follows
\begin{equation}
\partial_{x}q_{\rm o}(x)=
-\gamma(v)
\left(\frac{\sigma_{0} }{Mc_{l}^{2} }\right) \Bigl( a_{l}^{3}\,\phi_{\rm o}^{2}(x)\,\Bigr)
\,+\, \gamma (v)\, \vert \kappa_{3} \vert \left(\gamma (v)\,
\frac{\sigma_{0} }{Mc_{l}^{2} }\right)^{2}
\Bigl(a_{l}^{3}\,\phi_{\rm o}^{2}(x)\,\Bigr)^{2}.
\label{displ1}
\end{equation}
To estimate its amplitude, $\partial_{x}q_{\rm o}$,
we have to estimate the parameter \,$a_{l}^{3}\,\Phi_{\rm o}^{2}$ first.

For $\bar{n}_{\rm o} > 10$ 
\,and\,  $\vert \tilde{\varepsilon_{0}}\,(v)\vert \simeq (10^{-1}\sim 1)\,{\rm Ry}^{*}$ 
(i.e. \,$\gamma(v) \ge 2\gamma_{\rm o}$),
we obtain
$$
a_{l}^{3}\,\Phi_{\rm o}^{2} \simeq \frac{ a_{l}^{3} }{ \tilde{a}_{B}^{3}   }
\,\frac{ \vert \tilde{\varepsilon}_{0}(v)\vert   }{x\,{\rm Ry}^{*}}
\,\frac{1}{2\,\bar{n}_{\rm o}^{2}} \,\sim\, 
\frac{ a_{l}^{3} }{ \tilde{a}_{B}^{3}   }\,\frac{1}{2\,\bar{n}_{\rm o}^{2}}
\,\propto \,N_{\rm o}^{2}.
$$
If this parameter is small enough,  such as 
$a_{l}^{3}\,\Phi_{\rm o}^{2}(N_{\rm o},\,v) \simeq 10^{-3}\sim 10^{-5}$, 
\,
we can neglect the nonlinear corrections to the amplitude $\partial_{x}q_{\rm o} < 0$
and to the shape of $\partial_{x}q_{\rm o}(x)$ as well, 
\begin{equation}
\partial_{x}q_{\rm o}=-\gamma(v)\,\frac{\sigma_{0} }{Mc_{l}^{2} }\,\Bigl( a_{l}^{3}\,\Phi_{\rm o}^{2}\Bigr)\,
\Bigl\{1\,- \,\gamma (v)\, \vert \kappa_{3} \vert 
\left(\gamma (v)\,\frac{\sigma_{0} }{Mc_{l}^{2} }\right)
\Bigl(a_{l}^{3}\,\Phi_{\rm o}^{2}\Bigr)\,\Bigr\} \approx 
-\gamma(v)\Bigl(\frac{\sigma_{0} }{Mc_{l}^{2} }\Bigr) \bigl( a_{l}^{3}\,\Phi_{\rm o}^{2}\bigr).
\label{dispAMP}
\end{equation}
Thus, due to the validity of \,$n_{\rm o}\,\tilde{a}_{B}^{3} \ll 1$, there is almost no difference
between the approximation  
\begin{equation}
\phi_{\rm o}(x) \cdot \partial_{x}q_{\rm o}(x) \approx \Phi_{\rm o}\,
 \cosh^{-1}(\beta(\Phi_{\rm o})\,x) \cdot 
\bigl( - \vert \partial_{x}q_{\rm o} \vert\, \bigr)\, \cosh^{-2}(\beta(\Phi_{\rm o})\,x),  
\end{equation}
where  
\begin{equation}
\beta(\Phi_{\rm o}) \equiv L_{0}^{-1} \approx 
\frac{ \vert \tilde{\varepsilon}_{0}(v)\vert   }{x\,{\rm Ry}^{*}}\, \frac{1}{2 \, \bar{n}_{\rm o} \,\tilde{a}_{B} },
\,\,\,\,\,\,\,N_{\rm o}= N_{\rm o}^{*}/\bar{n}_{\rm o},
\,\,\,\,\,\,\bar{n}_{\rm o} \gg 1,
\label{need1}
\end{equation}
and the exact solution of the  weakly nonlinear case with $\nu_{1} >0$ and $\kappa_{3}<0$.
For  $S\simeq (10^{-2} \sim 10^{-3})$\,cm$^{2}$, $ \tilde{a}_{B}^{2} \simeq (25 \sim 50)\,10^{-16}$\,cm$^{2}$, 
we estimate  $ N_{\rm o}^{*} \simeq 10^{13} \sim 10^{14} $. Although 
the approximate solutions we used in this study are valid
for\,  $ N_{\rm o} \ll  N_{\rm o}^{*}$, they can be used at
\,$ N_{\rm o} <  N_{\rm o}^{*}$ for estimates.

Note that  the effective chemical potential is a rather small parameter in this model,
\begin{equation}
\vert \tilde{\mu} \vert(N_{\rm o},\,v) \approx \frac{ \vert \widetilde{\nu_{0}}\,(v)\vert ^{2}}{
4\,\bar{n}_{\rm o}^{2}\,x\,{\rm Ry}^{*}\,\tilde{a}_{B}^{6} } \approx 
\frac{ \vert \tilde{\varepsilon}_{0}(v)  \vert  }{4\,\bar{n}_{\rm o}^{2} }\,
\left( \frac{ \vert \tilde{\varepsilon}_{0}(v)  \vert  }{x\,{\rm Ry}^{*} }     \right). 
\label{need2}
\end{equation}
That is why the characteristic length,  $L_{\rm ch} \propto \vert \tilde{\mu} \vert^{-1/2}$, see (\ref{Width}), 
can be estimated as $(10^{2}\sim 10^{4})\,\tilde{a}_{B}$ within the validity of approximations 
(\ref{AMP1}), (\ref{MU1}).
Moreover,  $\vert \tilde{\mu} \vert/ \mu^{*} \le 10^{-2} $ and
the parameter \,$\eta_{1}(\Phi_{\rm o})$ in (\ref{GENform})
can be estimated as $\sim 10^{-2}$.  In this case, one can 
neglect it in Eq.  (\ref{SolitonMod}). 

Returning to the laboratory reference frame, we can write the condensate wave function 
in the form (see Fig. 1):  
$$ 
\psi_{0}(x,t)\cdot u_{0}(x,t)\delta_{1j} \approx
\exp\left(-i\left(\tilde{E}_{g}+\frac{mv^{2}}{2} -\vert\tilde{\mu}\vert \right )t \,\right)
\exp\bigl(i(\varphi + mvx)\,\bigr)\times
$$
\begin{equation}
\times \Phi_{\rm o}\cosh^{-1}\bigl( L_{0}^{-1}(x-vt)\,\bigr )\cdot
\Bigl( Q_{\rm o}\,-\,Q_{\rm o}\tanh \bigl(L_{0}^{-1}(x-vt)\bigr)\, \Bigr ),
\label{movingcond}
\end{equation}
where we count the exciton energy from the bottom of the crystal valence band, ($\tilde{E}_{g}< E_{\rm gap}$),  
and  $2Q_{\rm o}(N_{\rm o},v)$ is the amplitude of the phonon part of 
condensate,
$$
Q_{\rm o} \approx  
\gamma(v)\,\Bigl(\frac{\sigma_{0} }{Mc_{l}^{2} }\Bigr)\,\left(
\frac{ a_{l}^{2}}{\tilde{a}_{B}^{2}}\, \frac{1}{\bar n_{\rm o}}\right )\,
a_{l} \,\ll\, a_{l}.
$$

To calculate the energy of the moving condensate within the Lagrangian approach, 
(see Eq. (\ref{lagrangian})\,), we have to integrate the zeroth component
of the energy-momentum tensor ${\cal{T}}_{0}^{0}$ over the spatial coordinates.
Consequently, we have the 
following formula  
$$
{\cal{T}}_{0}^{0}({\bf x},t)= \tilde{E}_{g}\phi^{*}_{\rm o} \phi_{\rm o} + 
\frac{\hbar^{2}}{2m}\nabla\phi^{*}_{\rm o}\,\nabla\phi_{\rm o} 
+ \frac{\nu_{0}}{2}\left( \phi^{*}_{\rm o} \right)^{2}\phi_{\rm o}^{2}
+\frac{\nu_{1}}{3}\left( \phi^{*}_{\rm o} \right)^{3}\phi_{\rm o}^{3}
+
$$
$$
+ \,\frac{\rho}{2}(\partial_{t} q_{\rm o})^{2} +\frac{\rho c_{l}^{2}}{2}\,(\partial_{x} q_{\rm o})^{2}
+ \frac{\rho c_{l}^{2}}{2}\, \kappa_{3}\, (\partial_{x} q_{\rm o})^{3}
+
\sigma_{0}\,\phi^{*}_{\rm o}\phi_{\rm o}\,\partial_{x} q_{\rm o}. 
$$
Here we do not take into account a small correction to this energy due to the quantum 
depletion of the condensate ($\langle \delta \Psi ^{\dag} \delta \Psi ({\bf x}) \rangle _{T=0}\, \ne 0$
\,and \,$\langle ( \partial_{x}\delta u_{j})^{2} \rangle _{T=0} \,\ne 0$).
Then the result reads 
$$
E _{\rm o}(N_{\rm o},\,v)=\int\!d{\bf x}\,{\cal{T}}_{0}^{0} = E_{\rm ex} + E_{\rm int} + E_{\rm ph}\approx
$$
\begin{equation}
\approx N_{\rm o}\left(\tilde{E}_{g} + \frac{m\,v^{2}}{2}\,\right) - 
N_{\rm o}\left(\vert\tilde{\mu}\vert + \nu_{0}\,\Phi_{\rm o}^{2}/3
\,\right)
+  N_{\rm o} \left\{ \frac{M\,(c_{l}^{2}+v^{2})}{2}\,\gamma^{2}(v)
\,\left( \frac{\sigma_{0}}{M c_{l}^{2}} \right)^{2}\,
\frac{2}{3}\, \bigl( a_{l}^{3}\,\Phi_{\rm o}^{2} \bigr)\, \right\}.
\label{Energy}
\end{equation}
We  will disregard  the terms  $\sim N_{\rm o}\,\nu_{1}\Phi_{\rm o}^{4}$  in  \,$E_{\rm int} <0$,   
and the corrections $\propto \vert \kappa_{3} \vert $ in  \,$E_{\rm ph}$.
Then we can write  
\begin{equation}
\vert E_{\rm int}  \vert / N_{\rm o}  \approx 
 \vert \widetilde{\nu_{0}} \,(v)\vert \, \Phi_{\rm o}^{2}/ 2
\,+\, \nu_{0}\,\Phi_{\rm o}^{2}/3 \,\simeq\,
\bigl(\nu_{0}/\tilde{a}_{B}^{3}\,\bigr)\,\Bigl(\tilde{a}_{B}^{3} \Phi_{\rm o}^{2}\,\Bigr) <  {\rm Ry}^{*}
\end{equation}
and 
\begin{equation}
E_{\rm ph}/ N_{\rm o} 
\approx \frac{M\,(c_{l}^{2}+v^{2})}{2} \,\vartheta(N_{\rm o},\,v)\,\simeq \,
M c_{l}^{2}\,\vartheta(N_{\rm o},\,v), 
\end{equation}
where
\begin{equation}
\vartheta(N_{\rm o},\,v) =
\left( \gamma (v)\,\frac{\sigma_{0}}{M c_{l}^{2}} \right)^{2}\,
\frac{2}{3}\, \bigl( a_{l}^{3}\,\Phi_{\rm o}^{2} \bigr) \ll 1.
\label{theta}
\end{equation}
Note that the parameter  $\vartheta(N_{\rm o},\,v) $ is a rather small one, 
\,$\vartheta \sim 
a_{l}^{3}\,\Phi_{\rm o}^{2} \simeq 10^{-3} \sim 10^{-5} $, 
so that the value of $E_{\rm ph}/ N_{\rm o}$  can be \,
$ <  {\rm Ry}^{*}$, and, roughly, \,$\Phi_{\rm o}^{2} \propto N_{\rm o}^{2}$.

One can see that the exciton-phonon condensate
carries a non-zeroth momentum,
\newline
$P_{{\rm o}\,x }= P_{{\rm ex},\,x}+ P_{{\rm ph},\,x}$:
$$
P_{{\rm o}\,x}=
\int \!d{\bf x}\, (\hbar / 2i)(\phi_{0}^{*}(x,t)\,\partial_{x}\phi_{0}(x,t)-
\partial_{x}\phi_{0}^{*}(x,t)\,\phi_{0}(x,t)\,)\, - \,\rho\,\partial_{t}u_{0}(x,t)\,\partial_{x}u_{0}(x,t)=
$$
$$
=\int d{\bf x}\,mv\,\phi_{\rm o}^{2}(x)\, + \,\rho v\! \left ( \gamma(v)\,\frac{\sigma_{0} }{M c_{l}^{2} }\,
a_{l}^{3}\,\phi_{\rm o}^{2}(x)\,\right)^{2} \approx \, N_{\rm o}\,mv \,+\,N_{\rm o}\,Mv\,
\vartheta(N_{\rm o},\,v)
\,\equiv 
$$
\begin{equation}
\equiv\,
N_{\rm o} \,m\left\{1+  (M/m)\vartheta(N_{\rm o},\,v)\,\right\} v.
\end{equation}
Thus, we obtain \,$m_{\rm eff }=m\left\{1+  (M/m)\vartheta(N_{\rm o},\,v)\,\right\}$
and estimate  the parameter 
\newline
$(M/m)\,\vartheta(N_{\rm o},\,v) \simeq  1 \sim 5$ 
\,at \,$\gamma(v) \ge 2\gamma_{\rm o}$, \,$\bar{n}_{\rm o} \ge 10$.
 
\newpage
\begin{figure}
\begin{center}
\leavevmode
\epsfxsize = 400pt
\epsfysize = 300pt
\epsfbox{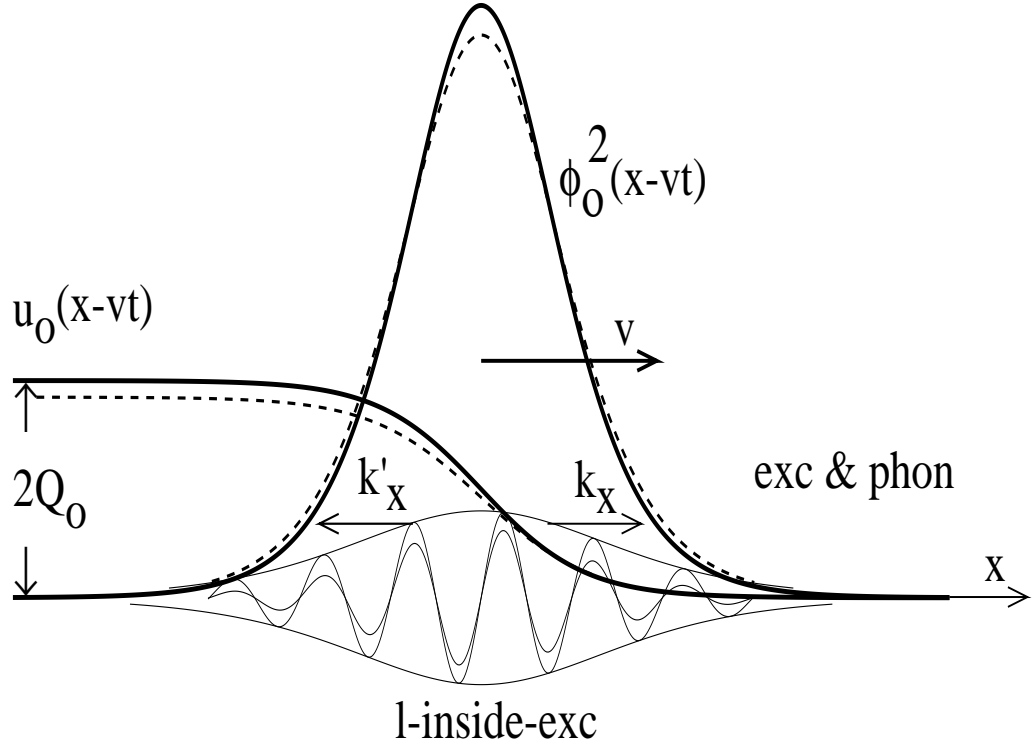}
\end{center}
\vspace*{1cm}
\caption{  
Moving exciton-phonon condensate, 
as it appears in the quasi-stationary model,
$\phi_{\rm o}(x-vt)\cdot u_{\rm o}(x-vt)\delta_{1j}$, 
is presented by bold lines on this Figure.
Here,
$2Q_{\rm o}$ is the amplitude of the coherent phonon 
state $u_{\rm o}(x-vt)$, and $\Phi_{\rm o}$
is the amplitude of the macroscopic wave function of excitons.
Longitudinal exciton-phonon excitations
(${\bf k}\,||\, Ox$) of the condensate 
are schematically depicted. Under transformation 
$N_{\rm o} \rightarrow N _{\rm o}   - \delta N$,
the condensate wave function  is changed as it is 
presented  by dashed lines. 
}
\label{l-inside}
\end{figure}

\section{Low-Lying Excitations of Exciton-Phonon Condensate}

To consider the stability of the exciton-phonon condensate 
moving in a lattice,  one has  to couple the excitons with 
different sources of perturbation, such as impurities,  
thermal lattice phonons, surfaces, etc..
In this work, however, we will not specify any source. 
Instead, we consider 
the stability conditions in relation to
creation (emission) of the condensate 
excitations that can be found 
in the framework of investigation of the low-energy
excitations of the condensate itself.    

Although the condensate wave function $\phi_{\rm o}(x)\cdot q_{\rm o}(x)$ was 
obtained 
in the framework of the effective 1D model, we normalized it in 3D space.
Therefore, we can use  
this solution as a classical part in the following  decomposition of 
3D field operators in the co-moving frame:    
\begin{equation}
 \hat{\psi}({\bf x},t)=\exp(-i\mu t)\,(\phi_{\rm o}(x)+ \delta\hat{\psi}
({\bf x},t)\,),
\label{hatpsi}
\end{equation}
\begin{equation}
\hat{u}_{j}({\bf x},t) =q_{\rm o}(x)\,\delta_{1j} + \delta\hat{u}_{j}({\bf 
x},t), 
\label{hatu}
\end{equation}
where $\mu=\tilde{\mu} - mv^{2}/2 $. \,Substituting 
the field operators of the form (\ref{hatpsi}),(\ref{hatu}) 
into the Lagrangian density (\ref{lagrangian}), we can write the 
later in the following form:
\begin{equation}
{\cal L}={\cal L}_{\rm o}({\rm e}^{-i\mu t}\phi_{\rm o}(x),\,q_{\rm 
o}(x)\delta_{1j}\,)\, +\, 
{\cal L}_{2}(\delta\hat{\psi}^{\dag}({\bf x},t),\, 
\delta\hat{\psi}({\bf x},t), \,\delta \hat{\bf u}({\bf x},t)\,)\,+... ,  
\label{lagrangian1}
\end{equation}  
where ${\cal L}_{\rm o}$ stands for the classical part of ${\cal L}$, and 
${\cal L}_{2}$ is the bilinear form in the $\delta$-operators.

In the simplest (Bogoliubov) approximation \cite{Fetter},\cite{Pit},
\,${\cal L}\approx {\cal 
L}_{\rm o}+ {\cal L}_{2}$ and, hence, the bilinear form ${\cal L}_{2}$ 
defines the equations of motion for the fluctuating parts of the field 
operators. 
  As a result, these equations are 
{\it linear} and can be written as follows:     
$$
i\hbar\partial_{t}\,\delta\hat{\psi}({\bf x},t)=
\left(-\frac{\hbar^{2}}{2m}\Delta + \vert\tilde{\mu}\vert +
\left\{ \nu_{0} +  \widetilde{\nu_{0}}(v)  \right \}
\phi_{\rm o}^{2}(x) 
+ \left\{ 2\nu_{1} +  \widetilde{\nu_{1}}(v)\right\}
\phi_{\rm o}^{4}(x)\,\right)
\delta\hat{\psi}({\bf x},t)\,+ 
$$
\begin{equation}
+ \left(\nu_{0}\phi_{\rm o}^{2}(x) + 2\nu_{1}\phi_{\rm o}^{4}(x)\,\right)\delta\hat{\psi}^{\dag}({\bf x},t)
+ \sigma_{0}\phi_{\rm o}(x)\nabla \delta\hat{\bf u}({\bf x},t),
\label{deltapsi}
\end{equation}
and 
$$
\left(
\partial_{t}^{2}-c_{l}^{2}\Delta - v(\partial_{t}\partial_{x} + \partial_{x} \partial_{t})
+v^{2}\partial_{x}^{2}\right)
\delta\hat{ u}_{j}({\bf x},t)\, =
$$
\begin{equation}
=\,
\rho^{-1}\sigma_{0}\partial_{j}\Bigl( \phi_{\rm o}(x)\bigl(\delta\hat{\psi}({\bf x},t) 
+\delta\hat{\psi}^{\dag}({\bf x},t)\bigr )\,\Bigr),\,\,\,j=2,3(\equiv \perp),
\label{deltau}
\end{equation}
$$
\left(\partial_{t}^{2}-c_{l}^{2}\Delta - v(\partial_{t}\partial_{x} + \partial_{x} \partial_{t})
+v^{2}\partial_{x}^{2}\right)
\delta\hat{ u}_{x}({\bf x},t)\,-
$$
$$
-  c_{l}^{2} 2\kappa_{3}\, (\partial_{x} q_{\rm o}(x) )\, \partial_{x}^{2}\delta\hat{ u}_{x}({\bf x},t)
-  c_{l}^{2} 2\kappa_{3}\, (\partial_{x}^{2} q_{\rm o}(x) )\,\partial_{x} \delta\hat{ u}_{x}({\bf x},t)=
$$
\begin{equation}
=\rho^{-1}\sigma_{0}\,\partial_{x}\!\left( \phi_{\rm o}(x)\left(\delta\hat{\psi}({\bf x},t) 
+\delta\hat{\psi}^{\dag}({\bf x},t)\right )\,\right), \,\,\,\,\,j=1(\equiv x).
\label{deltau1}
\end{equation}

The same approximation can be performed within the Hamiltonian approach.
Indeed,  decomposition of the field operators near 
their nontrivial classical parts
leads to the decomposition of the Hamiltonian (\ref{ham1}) itself, and -- 
as it was done with the Lagrangian -- 
only the classical part of $\hat{H}$, $H_{\rm o}$, and  
the bilinear form in the fluctuating fields, $\hat{H}_{2}$, are left for examination: 
\begin{equation}
\hat{H} \approx  H_{\rm o}( \psi_{0}^{*}, \psi_{0}, \,\pi_{0}, u_{0})+
H_{2}(\delta\hat{\psi}^{\dag},\,\delta\hat{\psi}, \,\delta\hat{\pi}, 
\delta\hat{u}).
\label{ham2}
\end{equation} 

In the co-moving  frame, \,$\hat{\pi}_{j}=\rho\,\partial_{t}{\hat u}_{j}- \rho v\,\partial_{x}{\hat u}_{j}$, i.e.
$$
\pi_{\rm o\,j}(x)=- \rho v\, \partial_{x} q_{\rm o}(x)\,\delta_{1j} \ne 0\,\,\,\,\,{\rm  and}\,\,\,\,\, 
\delta\hat{\pi}_{j}=\rho\,\partial_{t}\delta{\hat u}_{j}- \rho v\,\partial_{x}\delta{\hat u}_{j}, 
$$
and the standard commutation relation, 
$[\, \delta{\hat u}_{j}({\bf x},t),\,\,\delta{\hat \pi}_{s}({\bf x}',t)\,]$,  has the form  
\begin{equation}
[\, \delta{\hat u}_{j}({\bf x},t),\,\,\rho\,\partial_{t}\delta{\hat u}_{s}({\bf x}',t)   - \rho  v\,\partial_{x}\delta{\hat u}_{s}({\bf x}',t)\,] 
= i\hbar\,\delta({\bf x}-{\bf x'})\,\delta_{js}.
\label{comrel1}
\end{equation}
However, the Hamiltonian (\ref{ham2}) can be diagonalized
and rewritten in the form:  
\begin{equation}
\hat{H}=H_{\rm o}({\rm e}^{-i\mu t}\phi_{\rm o}(x),\,q_{\rm o}(x)\,)
+\delta E_{\rm o} + 
\sum_{1,s}\hbar\omega_{1,s}\, \hat{\alpha}^{\dag}_{1,s}\hat{\alpha}_{1,s}
+ \sum_{2,s}\hbar\omega_{2,s}\, \hat{\alpha}^{\dag}_{2,s}\hat{\alpha}_{2,s}.
\label{ham3}
\end{equation}
Here, $\delta E_{\rm o}$ is the quantum correction to the energy of
the condensate and the indexes ${1,s}$  and ${2,s}$ label the elementary excitations of the system. 
We assume the operators $\hat{\alpha}^{\dag}_{j,s}$, 
$\hat{\alpha}_{j,s}$ are the Bose ones. These operators describe  two different branches of the excitations,  $j=1,2$,
and they can be represented by the following 
linear combinations of the ``delta- 
operators'':
\begin{equation}
\hat{\alpha}_{j,s}=\int \!\!d{\bf x} \left(\,U_{j,s}({\bf x})\,\delta\hat{\psi}({\bf x})+
                 V_{j,s}({\bf x})\,\delta\hat{\psi}^{\dag}({\bf x}) 
+Y_{j,s}^{i}({\bf x})\,\delta\hat{u}_{i}({\bf x})
+Z_{j,s}^{i}({\bf x})\,\delta\hat{\pi}_{i}({\bf x})\,\right),
\label{alpha}
\end{equation}
\begin{equation}
\hat{\alpha}_{j,s}^{\dag}=\int\!\! d{\bf x} \left(\,U_{j,s}^{*}({\bf x})\,\delta\hat{\psi}^{\dag}({\bf x})+
                 V_{j,s}^{*}({\bf x})\,\delta\hat{\psi}({\bf x}) 
+Y_{j,s}^{i^{*}}({\bf x})\,\delta\hat{u}_{i}({\bf x})
+Z_{j,s}^{i^{*}}({\bf x})\,\delta\hat{\pi}_{i}({\bf x})\,\right).
\label{alpha1}
\end{equation}
Note that 
by analogy with the exciton-polariton modes in semiconductors\,\cite{Hopfield},\cite{Keldysh1}
the excitations 
of the condensate (\ref{movingcond}) can be considered as a mixture of 
exciton- and phonon-type modes. However, in this model, the phonons are  fluctuations
of the \,$\bigl(\pi_{0}( x,t),\,u_{0}(x,t)\,\bigr)$\,-\,part of the condensate.  
The commutation relations between $\alpha$-operators are Bose ones, so that  
$$
[\, \hat{\alpha}_{1,s},\,\, \hat{\alpha}_{1,s'}^{\dag}     \,]= \delta_{ss'}
$$
lead to the following orthogonality condition
$$
 \int \!\!d{\bf x}\,\Bigl( U_{1,s}U_{1,s'}^{*}({\bf x}) -  V_{1,s}V_{1,s'}^{*}({\bf x})\,\Bigr)
+ (i \hbar)\! \sum_{r=1,2,3}
\int \!\!d{\bf x}\,\Bigl(Y_{1,s}^{r}Z_{1,s'}^{r^{*}}({\bf x}) - Z_{1,s}^{r}Y_{1,s'}^{r^{*}}({\bf x})\,
\Bigr)
=\delta_{ss'}.
$$

Since the $\alpha$-operators (see Eq. (\ref{ham3})\,) evolve in time as simply as    
$$
\hat{\alpha}_{j,s}(t)={\rm e}^{-i\omega_{j,s}t}\,\hat{\alpha}_{j,s},\,\,\,\,\,\,
\hat{\alpha}_{j,s}^{\dag}(t) = 
{\rm e}^{i\omega_{j,s}t}\,\hat{\alpha}_{j,s}^{\dag},
$$
these operators (and the frequencies $\{ \omega_{j,s} \}$) are 
the eigenvectors (and, correspondingly, the eigenvalues) of the equations of 
motion (\ref{deltapsi}),(\ref{deltau}) obtained within the 
Lagrangian method.  
Then, the time dependent ``$\delta$-operators'' in 
 (\ref{deltapsi}),(\ref{deltau}) can be 
represented by the following linear combinations of the $\alpha$-operators: 
$$
\delta\hat{\psi}({\bf x},t)=\sum_{1,s} {\rm u}_{1,s}({\bf x})
                             \,\hat{\alpha}_{1,s}{\rm e}^{-i\omega_{1,s}t}
\,+\,{\rm v}_{1,s}^{*}({\bf x})\,\hat{\alpha}_{1,s}^{\dag}{\rm e}^{i\omega_{1,s}t}\,+
$$
\begin{equation}
\,+\, \sum_{2,s}     {\rm u}_{2,s}({\bf x})
                             \,\hat{\alpha}_{2,s}{\rm e}^{-i\omega_{2,s}t}
\,+\,{\rm v}_{2,s}^{*}({\bf x})\,\hat{\alpha}_{2,s}^{\dag}{\rm e}^{i\omega_{2,s}t},
\label{uvtransform1}
\end{equation}
$$
\delta\hat{u}_{r}({\bf x},t)=\sum_{1,s}C_{1,s}^{r}({\bf x})\,\hat{\alpha}_{1,s}
{\rm e}^{-i\omega_{1,s}t}+ C_{1,s}^{r ^{*}}({\bf x})\,\hat{\alpha}^{\dag}_{1,s}{\rm e}^{i\omega_{1,s}t}\,+\,
$$
\begin{equation}
+\,\sum_{2,s}C_{2,s}^{r}({\bf x})\,\hat{\alpha}_{2,s}
{\rm e}^{-i\omega_{2,s}t}+ C_{2,s}^{r^{*}}({\bf x})\,\hat{\alpha}^{\dag}_{2,s}{\rm e}^{i\omega_{2,s}t},
\label{uvtransform2}
\end{equation}
For $\delta\hat{\pi}_{r}({\bf x},t)$, one has to change $C_{j,s}^{r}({\bf x})$ \,to\, 
$D_{j,s}^{r}=\rho \,(-i\omega_{j,s} -  v\partial_{x})C_{j,s}^{r}({\bf x})$
in (\ref{uvtransform2}).
Note that this {\it ansatz} 
is, in fact, a generalization of the u-v Bogoliubov
transformation. 

Then we can rewrite Eqs. (\ref{alpha}),(\ref{alpha1}) as follows ($j=1,2$)
\begin{equation}
\hat{\alpha}_{j,s}=\int \!\!d{\bf x} \left(\,{\rm u}_{j,s}^{*}({\bf x})\,\delta\hat{\psi}({\bf x})-
                 {\rm v}_{j,s}^{*}({\bf x})\,\delta\hat{\psi}^{\dag}({\bf x}) 
- (i/\hbar)D_{j,s}^{r^*}({\bf x})\,\delta\hat{u}_{r}({\bf x})
+(i/\hbar) C_{j,s}^{r^*}({\bf x})\,\delta\hat{\pi}_{r}({\bf x})\,\right),
\label{alphaUV}
\end{equation}
\begin{equation}
\hat{\alpha}_{j,s}^{\dag}=\int\!\! d{\bf x} \left(\,{\rm u}_{j,s}({\bf x})\,\delta\hat{\psi}^{\dag}({\bf x}) -
                 {\rm v}_{j,s}({\bf x})\,\delta\hat{\psi}({\bf x}) 
+(i/\hbar) D_{j,s}^{r}({\bf x})\,\delta\hat{u}_{r}({\bf x})
- (i/\hbar)C_{j,s}^{r}({\bf x})\,\delta\hat{\pi}_{r}({\bf x})\,\right),
\label{alphaUV1}
\end{equation}
and one of the orthogonality relations has the form ($s=s'$)
$$
\int\!d{\bf x}\,\bigl( \vert {\rm u}_{1,s}({\bf x})\vert ^{2} - 
\vert {\rm v}_{1,s}({\bf x}) \vert ^{2}\,\bigr) \,+\,
$$
\begin{equation}
+\,(i/\hbar)\!\sum_{r=1,2,3} \!\int\!d{\bf x}\, 
\left( 
C_{1,s}^{r^*}\,\rho (-i\omega_{1,s} -  v\partial_{x}) C_{1,s}^{r} ({\bf x}) \,+\,
\rho (-i\omega_{1,s} + v\partial_{x}) C_{1,s}^{r^ *}\,C_{1,s}^{r }({\bf x})\,  \right) = 1.
\label{impORT22}
\end{equation}

The question we want to clarify is whether coupling between  excitonic excitations 
and phonon excitations is important for understanding   the condensate excitations.
Substituting {\it ansatz} (\ref{uvtransform1})-(\ref{uvtransform2})
into Eqs. (\ref{deltapsi}),(\ref{deltau}), 
we obtain the following coupled eigenvalue equations \cite{Loutsenko}:  
\begin{equation}
(\hat{L}(\Delta) - \hbar\omega_{j,s})\,{\rm u}_{j,s}({\bf x}) + \left(\nu_{0}\phi_{\rm o}^{2}( x) + 
2\nu_{1}\phi_{\rm o}^{4}( x)\right)
{\rm v}_{j,s}({\bf x}) + \sigma_{0}\phi_{\rm o}( x)\partial_{r} { C}_{j,s}^{r}({\bf x})=0,
\label{eigfunction1}
\end{equation}
\begin{equation}
\left(\nu_{0}\phi_{\rm o}^{2}( x) + 
2\nu_{1}\phi_{\rm o}^{4}( x)\right){\rm u}_{j,s}({\bf x})+
(\hat{L}(\Delta) +\hbar\omega_{j,s})\,{\rm v}_{j,s}({\bf x}) +
 \sigma_{0}\phi_{\rm o}( x)\partial_{r}{ C}_{j,s}^{r}({\bf x})=0,
\label{eigfunction2}
\end{equation}
\vspace*{0.1cm}
$$
-\rho^{-1}\sigma_{0}\,\partial_{r}\Bigl(\phi_{\rm o}( x)\,{\rm u}_{j,s}({\bf x})\Bigr)
-\rho^{-1}\sigma_{0}\,\partial_{r} \Bigl(\phi_{\rm o}( x)\,{\rm v}_{j,s}({\bf x})\Bigr)\,+
$$
\begin{equation}
+\,\left[(-i\omega_{j,s}- v\partial_{x})^{2} - c_{l}^{2}\Delta \right]{C}_{j,s}^{r}({\bf x})=0,
\,\,\,\,r=2,3,
\label{eigfunction3} 
\end{equation}
\vspace*{0.1cm}
$$
-\rho^{-1}\sigma_{0}\,\partial_{x}\Bigl(\phi_{\rm o}( x)\,{\rm u}_{j,s}({\bf x})\Bigr)
-\rho^{-1}\sigma_{0}\,\partial_{x} \Bigl(\phi_{\rm o}( x)\,{\rm v}_{j,s}({\bf x})\Bigr)\,+
$$
\begin{equation}
+\,
\left[(-i\omega_{j,s}-v\partial_{x})^{2} \,-\, c_{l}^{2}
\bigl(1+\vert \kappa_{3}\vert F_{3}(x)\bigr )\,\partial_{x}^{2}
\,-\, c_{l}^{2} \partial_{\perp}^{2} 
\,- \,c_{l}^{2} \vert \kappa_{3}\vert\bigl(\partial_{x} F_{3}(x)\bigr)
\,\partial_{x}\, 
\right]{ C}_{j,s}^{x}({\bf x})=0.
\label{eigfunction4} 
\end{equation}
Here we used the following notations 
$$
\hat{L}(\Delta)= (-\hbar^{2}/2m)\Delta +\vert\tilde{\mu }\vert + 
\bigl\{\nu_{0} +\widetilde{\nu_{0}}\,(v)\,\bigr\}\phi_{\rm o}^{2}( x)+
\bigl\{ 2\nu_{1} + \widetilde{\nu_{1}}\,(v)\bigr\}\phi_{\rm o}^{4}(x),
$$
$$
F_{3}(x)= 2\,\gamma(v) \Bigl(\sigma_{0}/M c_{l}^{2}\Bigr)\,a_{l}^{3}
\phi_{\rm o}^{2}( x).
$$

To simplify investigation of the characteristic properties of the different 
solutions of  Eqs. 
(\ref{eigfunction1})-(\ref{eigfunction3}),
we  subdivide the excitations (\ref{uvtransform1})-(\ref{uvtransform2}) into two major parts, the 
{\it inside}-excitations  and the {\it outside}-ones.
The {\it inside}-excitations
are  
localized merely inside the packet area, i.e. $\vert {\bf x} \vert < 2L_{0}$ and 
$\phi_{\rm o}^{2}(x)\approx \Phi_{\rm o}^{2}$,  
whereas the {\it outside}-excitations
propagate merely in the outside area, i.e. $\vert {\bf x}\vert > 2L_{0}$
and $\phi_{\rm o}^{2}(x) \simeq 4 \Phi_{\rm o}^{2}\,\exp(-2\vert x \vert /L_{0})
\rightarrow 0$. 

\subsection{Outside-Excitations}

For the outside collective excitations, the asymptotics of the low-lying  
energy spectrum can be found easily.  Indeed, if we assume that       
$\phi_{\rm o}^{2}(x) \approx 0$ 
and $\partial_{x}q_{\rm o}(x) \approx 0$
in the outside packet area, the equations 
(\ref{deltapsi})
and (\ref{deltau}) are (formally) uncoupled. 
Then, Eq. (\ref{deltapsi}) describes the excitonic branch  
of the outside-excitations with the following dispertion low in the co-moving frame:
\begin{equation}
\hbar\omega_{{\rm ex}}({\bf k})\approx \vert\tilde{\mu}\vert + (\hbar^{2}/2m)k^{2},\,\,\,\,\, 
({\rm u}_{\bf k}({\bf x})\approx u_{\bf k}\,{\rm e}^{i{\bf k}{\bf x}},\,\,\,
{\rm v}_{\bf k}({\bf x})\approx 0),
\label{gap1}
\end{equation}
and \,$\omega_{\rm ph}({\bf k}')=c_{l}\vert {\bf k}' \vert $ in 
the laboratory frame of reference. 

Then the exciton field operator, which describes the exciton condensate 
with {\it one}  long-wavelength outside-excitation,  
has the following form:
$$
\psi({\bf x}, t)\simeq 
\exp\bigl(-i(\tilde{E}_{g}+mv^{2}/2-\vert\tilde{\mu}\vert)t\,\bigr)
\exp\bigl(i(\varphi + mvx)\,\bigr)\,
\phi_{\rm o}(x-vt)\,+ 
$$
\begin{equation}
+\, \exp(-i(\tilde{E}_{g}+mv^{2}/2-\vert\tilde{\mu}\vert)t\,)
\exp\bigl(i(\varphi + mvx)\,\bigr)
\left\{ \exp(-i(\vert\tilde{\mu}\vert+\hbar {\bf k}^{2}/2m + k_{x}v)t\,)\,
u_{\bf k} \exp(i{\bf k}{\bf x})\,\right\}.
\label{OUTside}
\end{equation}
It is easy to see that such a  collective  excitation,\, 
$$
\hbar\omega_{\rm ex}({\bf k}) =\vert\tilde{\mu}\vert + \hbar^{2} {\bf k}^{2}/2m +  \hbar k_{x}v,
$$ 
can be interpreted 
as a free exciton  with the energy  and the (quasi)momentum
$$
\varepsilon_{\rm x}(\tilde{k}) = \tilde{E}_{g}+ \hbar^{2}\tilde{\bf k}^{2}/2m 
\,\,\,\,\,{\rm and}\,\,\,\,\, 
\hbar\tilde{k}_{j}= \hbar k_{j} + mv\,\delta_{1j}.   
$$
Note that  the  condition $\hbar\omega_{\rm ex }> 0$ can be violated 
at the velocities close to $v_{\rm o}$, whereas \,$\varepsilon_{\rm x}$
is always positive.
Then the question is whether $\hbar\omega_{\rm ex } < 0$ 
 really means the condensate instability in relation to the creation of 
outside excitations. For example, being unstable, the condensate could continuously
emit outside-excitations, which form a sort of `tail' behind the localized packet.

Recall that the particle number $\hat{N}$ is not conserved in quantum  states with a condensate, 
and $\langle \delta N ^{2}\rangle \simeq N_{\rm o}$. 
However, 
for $N_{\rm o} \ge 10^{10}$ and $T\ll T_{c}$, the following estimate is valid 
$$
\sqrt{\langle \delta N ^{2}\rangle}/N  \simeq 1/\sqrt{ N_{\rm o} } \le 10^{-5}. 
$$
Therefore, we can compare the condensate energy $E_{\rm o}(N_{\rm o},\,v)$  and 
the energy of the condensate that emits excitons, or, equivalently,  the condensate with
outside excitations, $\langle u_{\bf k} \rangle \sim \sqrt{ \delta N }$. For simplicity's sake,
we consider \,$\delta N$ \,different wave vectors, $\{\tilde{\bf k}_{j}\}$,  to be close to each other, 
so that the values of $\langle \tilde{k}\rangle ^{2}$ \,and\, $\langle \tilde{k}_{x}\rangle$ 
are well-defined. (This is a  model of how the instability tail could be formed.)
We obtain (see Eqs. (\ref{Energy})-(\ref{theta}))
$$
E_{\rm o}(N_{\rm o}-\delta N,\,v) +  E_{\rm x}(\langle \tilde{k} \rangle,\delta N ) + 
E_{\rm ph}(\langle k' \rangle,\,\delta N)
\approx
$$
$$
\approx E_{\rm o}(N_{\rm o},\,v) +
3 \left(\vert\tilde{\mu}\vert + \frac {\nu_{0} \,\Phi^{2}_{\rm o} }{3}\,\right)\!\delta N 
-  3\,\frac{M\,(c_{l}^{2}+v^{2})}{2}\,\vartheta(N_{\rm o},\,v)\,\delta N \,+
$$
\begin{equation}
+\, \left(\,\frac{\hbar^{2} \langle \tilde{k}\rangle ^{2}}{2m} - \frac{ mv^{2} }{ 2}\,\right)\!\delta N  +  
\hbar c_{l}\vert\langle  k'\rangle  \vert \,\delta N.
\label{cond_and_exc1}
\end{equation}
For the momentum of the moving condensate with the  outside excitations, we have 
$$
P_{{\rm o}\,x}(N_{\rm o}-\delta N,\,v) +  \hbar \langle  \tilde{k}_{x}\rangle   \delta N + 
\hbar \langle   k' _{x} \rangle  \delta N 
 \approx 
$$
\begin{equation}
\approx 
P_{{\rm o}\,x}(N_{\rm o},\,v) + (\hbar \langle \tilde{k}_{x}\rangle - mv)\,\delta N  + 
\bigl(\hbar \langle  k' _{x}\rangle   - 3\,Mv\,\vartheta(N_{\rm o},\,v)\bigr)\,\delta N . 
\label{cond-and-exc1}
\end{equation}

Note that the energy and the momentum of the  phonon part of the condensate change after exciton emission. 
We hypothesize that the transformation
\,$N_{\rm o}\rightarrow N_{\rm o}- \delta N$ (with the emission of  outside excitons, see Figs. 1,2)
corresponds to the case in which    
the outside exciton and  the outside acoustic phonon appear {\it together}. Indeed,   
in the $k\rightarrow 0$  limit (i.e. \,$\lambda= 2\pi/k \gg L_{0}$), 
we approximately considered  the condensate collective excitations  
as being uncoupled.  However, 
the phonon $\hbar {\bf k}' $  can be emitted with the energy compensating 
the changement of \,$\delta E_{\rm ph}= - (3/2)\,M(c_{l}^{2}+v^{2})\,\vartheta(N_{\rm o},\,v)$ in the phonon part of 
the condensate energy. Moreover,     
the order of value of $\vert \delta E_{\rm ph} \vert $ is typical for the low-energy acoustic phonons,
$\sim  1$\,meV.
If \,$\hbar k'_{x}>0$, the emitted phonon can compensate the changement of  \,
$\delta P_{{\rm ph}\,x} = - 3\,Mv\,\vartheta(v)$ \,as well.

\begin{figure}
\begin{center}
\leavevmode
\epsfxsize = 400pt
\epsfysize = 300pt
\epsfbox{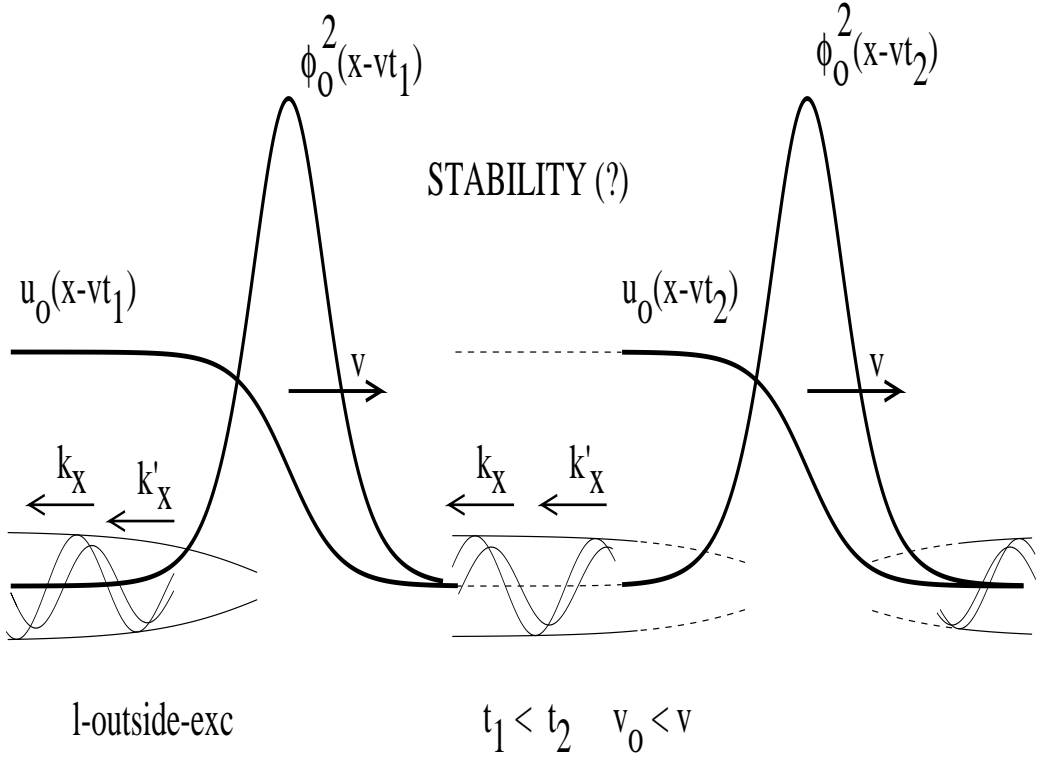}
\end{center}
\caption{
The ballistic condensate,
 $\phi_{\rm o}(x-vt)\cdot u_{\rm o}(x-vt)\delta_{1j}$,
seems to be stable in relation to 
emission of the 
 outside exciton-phonon excitations. 
(We consider the backward emission in the long-wavelength limit.)
The outside-excitations presented on this figure are  labeled by the wave vectors,   
$k_{x},\,\,k'_{x} < 0$ in 
the co-moving  frame. In the first approximation, the outside-excitations 
can be described  in terms of free excitons and free (acoustic) phonons emitted from
the condensate coherently.
}
\label{l-outside}
\end{figure}

The most interesting case is the backward emission of excitons, i.e.  
\,$\hbar k_{j} = \hbar k\,\delta_{1j} < 0 $ in the co-moving frame.
Then we can rewrite  (\ref{cond_and_exc1}) as follows
$$
E_{\rm o}(N_{\rm o}-\delta N,\,v) +  \langle \hbar\omega_{\rm ex}(\tilde{k})\rangle\,\delta N
+ \langle \hbar\omega _{\rm ph}( k')\rangle\,\delta N  \approx
$$
\begin{equation}
\approx
E_{\rm o}(N_{\rm o},\,v) +  \left(2\vert\tilde{\mu}\vert + \nu_{0} \,\Phi^{2}_{\rm o} \,\right)\!\delta N
+ 
\Bigl\{ \vert\tilde{\mu}\vert 
+ \frac{ \hbar^{2} \langle k\rangle ^{2}} {2m} - 
\hbar \vert \langle k \rangle \vert \,v \Bigr\}\delta N.
\label{Cond_And_Exc1} 
\end{equation}
The moving condensate can be considered as a stable one in relation to emission of the outside excitations  
($\delta N >\sqrt{\langle \delta N ^{2}\rangle}$)
if such an emission gains energy,
$$
E_{\rm o}(N_{\rm o}-\delta N,\,v)  + E_{\rm exc}(\langle\tilde{k}\rangle,\,\delta N)+ 
E_{\rm ph}(\langle k'\rangle,\,\delta N) > E_{\rm o}(N_{\rm o},\,v).
$$
This means that the following inequality has to be valid
\begin{equation}
\left\{ \vert\tilde{\mu}\vert + 
\frac{ \hbar^{2}\langle k \rangle ^{2}}{2m} - 
\hbar \vert \langle k\rangle\vert \,v  \right\}
+ (2\vert\tilde{\mu}\vert + \nu_{0} \,\Phi^{2}_{\rm o} \,) >0. 
\label{stabOUT}
\end{equation}
This condition  
can be rewritten in the dimensionless form as follows:
\begin{equation}
\Bigl( 
3 + 2\,\frac{ \nu_{0}}{\vert \widetilde{\nu_{0}}\,(v) \vert } + 
\left(\frac{2\pi}{\langle z \rangle }\right)^{\!2}\,\Bigr)
\, - \, \frac{2\pi}{\langle z \rangle}\, \frac{v}{c_{l}}
\,\sqrt{ 
\frac{ (4m/M)\,Mc_{l}^{2}/2\,}{\vert\tilde{\mu}\vert(N_{\rm o},\,v)}
}
>0.
\label{Stable}
\end{equation}
We argue that, 
even for velocities close to $v_{\rm o}$ 
(where\,  
$\vert \widetilde{\nu_{0}}\,(v) \vert$  can be  $\sim 0.1\, \nu_{0}$,  and the instability could appear
as \,$\vert\tilde{\mu}\vert(N_{\rm o},\,v) + \hbar^{2}k_{x}^{2}/2m - \hbar\vert k_{x}\vert v < 0$\,),
inequality  (\ref{stabOUT}) seems to be always true in the long-wavelength approximation,  
\,$k = (2\pi/ z)\,L_{0}^{-1} \le 10^{-1}\,L_{0}^{-1}$. 
On Fig.~\ref{l-outside}, 
the stable ballistic condensate is shown with its long-wavelength outside-excitations.

Note that the stability against  
large-$k$ modes cannot be properly described within approximation  (\ref{gap1}),(\ref{OUTside}).
However, we can discuss this case within the inside-approximation.   

\newpage

\subsection{Inside-Excitations}

To simplify the calculation of {\it inside}-excitation spectrum  
 (see Eqs. (\ref{eigfunction1})-(\ref{eigfunction3})\,) 
we will use the semiclassical approximation \cite{Pit}.
In this approximation, the excitations can be labeled 
by the wave vector ${\bf k}$ in the co-moving frame, and the following
representation holds:
\begin{equation}  
{\rm u}_{j,s}({\bf x})= {\rm u}_{j,{\bf k}}({\bf x})\,{\rm e}^{ i\varphi_{\bf k}({\bf x})},\,\,\, 
{\rm v}_{j,s}({\bf x})= {\rm v}_{j,{\bf k}}({\bf x})\,{\rm e}^{i\varphi_{\bf k}({\bf x})},\,\,\,
 C_{j,s}^{r}({\bf x})= C_{j,{\bf k}}^{r}({\bf x})\,{\rm e}^{ i\varphi_{\bf k}({\bf x})},
\label{semiclassical}
\end{equation}
where the phase \,$\varphi_{\bf k}({\bf x})\approx \varphi_{\rm o}+ {\bf kx}$, and ${\rm u}_{\bf k}({\bf x})$,
${\rm v}_{\bf k}({\bf x})$, and $C_{{\bf k}}({\bf x})$ are assumed to be  smooth functions of ${\bf x}$ in the inside condensate
area.
 Notice that the ${\bf k}$- and ${\bf x}$-representations are mixed here. This means 
that  
the operator nature
of the fluctuating fields is actually dismissed within the semiclassical approximation.
Howevever,  the 
orthogonality 
relations among ${\rm u}_{j,s}$, \,${\rm v}_{j,s'}$,  and $C_{j,s}$, \,$C_{j,s'}^*$,
and, hence, 
among \,${\rm u}_{j,{\bf k}}$, \,${\rm v}_{j,{\bf k}'}$, and  $C_{j,{\bf k}}$,  \,$C_{j,{\bf k'}}^*$
come from the Bose commutation relations between 
the operators $\alpha_{j,s}$ and $\alpha^{\dag}_{j,s'}$ \cite{Fetter},\cite{Pit}.  
For example, Eq. (\ref{uvtransform1}) is modified as follows:    
\begin{equation}
 \delta{\psi}({\bf x},t)\simeq \!\int\!\frac{d{\bf k}}{(2\pi)^{3}} \,
{\rm u}_{\bf k}({\bf x}){\rm e}^{i\varphi_{\bf k}({\bf x})}
                            {\rm e}^{-i\omega_{\bf k}({\bf x})t}
\,+\,{\rm v}_{\bf k}^{*}({\bf x}){\rm e}^{-i\varphi_{\bf k}({\bf x})}{\rm e}^{ i\omega_{\bf k}({\bf x})t},
\label{uvtransform111}
\end{equation}
and the {\it inside}-excitation part  of the elementary excitation term in 
Eq. (\ref{ham3}),
\newline
$\sum_{j,s}\cdots \approx \sum_{j,s,\,{\rm out}}+ \sum_{j,s,\,{\rm surf}} +\sum_{j,s,\,{\rm in}} \cdots $, 
can be written as  
\begin{equation}
\sum_{1,s,\,{\rm in}}\hbar\omega_{1,s}\hat{\alpha}_{1,s}^{\dag}\hat{\alpha}_{1,s}\simeq
\int\!\frac{d{\bf k}\,d{\bf x}}{(2\pi)^{3}}\,\hbar\omega_{1,{\bf k}}({\bf x})\,n_{1,{\bf k}}({\bf x}). 
\label{semienergy}
\end{equation}
Note that the semiclassical energy $\hbar\omega_{j,{\bf k}}(x)$
of the  inside-excitation 
mode $j,{\bf k}$ 
is supposed to be a smooth function of $x$ as well, (at least, as smooth as 
$\phi_{\rm o}^{2}(x)$, which is taken constant 
in the  inside-approximation). 

Although the low-lying excitations cannot be properly described within the semiclassical approximation, 
we apply it here to calculate the low-energy asymptotics 
of the spectrum. In fact, within the approximation $\bar{H} \approx H_{\rm o} + \bar{H}_{2}$, 
all the important properties of such excitations can be understood by use of 
the semiclassical  approach. 

There are two different types of the inside-excitations, the longitudinal
excitations and the transverse ones. The 
later have the wave vectors ${\bf k}$ perpendicular to the 
$x\,(v)$-direction. For the sake of simplicity, we choose \,${\bf k} \parallel Oy$. Then the vector  
$C_{j,{\bf k}}^{r}$ has one nonzeroth component for such transverse excitations, \,$C_{j,k}^{y} \ne 0$.     

Substituting {\it ansatz} (\ref{semiclassical}) with $ k_{r} = k_{\perp} \,\delta_{2,r}$ and $C_{j,{\bf k}}^{r}= C_{j,k}^{y}\,\delta_{2,r}$
into Eqs. (\ref{eigfunction1})-(\ref{eigfunction3}), 
we transform these differential equations into the algebraic ones 
(within the  inside-approximation \,$\hat{L}(\Delta) \rightarrow L(-{\bf k}^{2})$\,):
\begin{equation}
(L(- k_{\perp}^{2}) - \hbar\omega_{j,k})\,{\rm u}_{ k}({\bf x}) + \left(\nu_{0}\phi_{\rm o}^{2}({x}) + 
2\nu_{1}\phi_{\rm o}^{4}({ x})\right)
{\rm v}_{j,k}({\bf x}) +\sigma_{0}\phi_{\rm o}({x})\,i k_{\perp} { C}_{j,k}^{y}({\bf x})=0, 
\label{EIGfunction01}
\end{equation}
\begin{equation}
\left(\nu_{0}\phi_{\rm o}^{2}( x) + 
2\nu_{1}\phi_{\rm o}^{4}({ x})\right){\rm u}_{j, k}({\bf x})+
( L(- k_{\perp}^{2}) +\hbar\omega_{j,k})\,{\rm v}_{j,k}({\bf x}) +
 \sigma_{0}\phi_{\rm o}({ x})\,i{ k_{\perp}}{ C}_{j,k}^{y}({\bf x})=0
\label{EIGfunction02}
\end{equation}
\begin{equation}
\rho^{-1}\sigma_{0}
\phi_{\rm o}(x)\,i k_{\perp} \,{\rm u}_{j, k}({\bf x})
+\rho^{-1}\sigma_{0}
\phi_{\rm o}(x)\,ik_{\perp} \,{\rm v}_{j,k}({\bf x})+
\left[\,\omega_{j,k}^{2} - c_{l}^{2} k_{\perp}^{2} \right]
C_{j, k}^{y}({\bf  x})=0.
\label{EIG} 
\end{equation}  
After some straightforward algebra, we can write out the equation that defines the spectrum of  
transverse exciton-phonon excitations in the  inside-approximation:   
$$  
\left(\,\omega_{j,k}^{2} - c_{l}^{2}{ k_{\perp} }^{2}\right)\times
$$
$$
\times \left[\, (\hbar\omega_{j,k})^{2}-
\left( L(- k_{\perp}^{2})-
 \nu_{0}\phi_{\rm o}^{2}({ x}) - 
2\nu_{1}\phi_{\rm o}^{4}({ x})\,\right)
\left( L(- k_{\perp}^{2})+
 \nu_{0}\phi_{\rm o}^{2}({ x}) + 
2\nu_{1}\phi_{\rm o}^{4}({ x})\,\right)
\,\right]=
$$
\begin{equation}
=\left(\,
 L(- k_{\perp}^{2})-
 \nu_{0}\phi_{\rm o}^{2}({ x}) - 
2\nu_{1}\phi_{\rm o}^{4}({ x})
\,\right)\,
\,\frac{2\,\sigma_{0}^{2}}{\rho\,c_{l}^{2}}\,\phi_{\rm o}^{2}(x)\,(c_{l}^{2} k_{\perp}^{2}).
\label{SPECtrum0}
\end{equation}
Taking into account the momentum cut-off $k_{0}$,  which is defined as
$$
(\hbar^{2} /2m )\,k_{0}^{2} \approx  \vert \tilde{\mu} \vert = (\hbar^{2} /2m )\,L_{0}^{-2},
\,\,\,\,\,k>k_{0},
$$
 we can rewrite 
Eq. (\ref{SPECtrum0}) as follows
$$  
\left(\,\omega_{j,k}^{2} \,- \,c_{l}^{2} k_{\perp}^{2}\,\right)\times
$$
$$
\times \left[\, (\hbar\omega_{j,k})^{2}-
\left(
\frac{ \hbar^{2} }{2m} [k_{\perp}^{2}-{ k}_{0}^{2} ]+
F(x)+ \epsilon_{+}\right)\!\cdot\! 
\left( \frac{\hbar^{2} }{2m} [ k_{\perp}^{2}-{ k}_{0}^{2} ] +
F(x)+2\nu_{0}\phi^{2}_{\rm o}(x) + \epsilon_{+}'\right) \,\right]=
$$
\begin{equation}
=\left(\,
\frac{ \hbar^{2} }{2m} [ k_{\perp}^{2}-k_{0}^{2} ]+
F(x)+ \epsilon_{+}\right)\,2 \frac{\sigma_{0}}{M c_{l}^{2}}\,(\sigma_{0}
 a_{l}^{3})\,\Phi_{\rm o}^{2} \,(c_{l}^{2} k_{\perp}^{2}), \,\,\,\,\,\, \vert x \vert < L_{0}.
\label{spectrum0tt}
\end{equation}
Here \,$F(x)=\vert \widetilde{\nu_{0}}\,(v) \vert (\Phi_{\rm o}^{2} - \phi_{\rm o}^{2}(x)\,)>0$, 
i.e. \,$F(x)\simeq 0$ inside the condensate, and 
$$
 \epsilon_{+} \approx \widetilde{\nu_{1}}\, \phi_{\rm o}^{4}(x), \,\,\,\,\,\,
 \epsilon_{+}' \approx 5 {\nu_{1}}\, \phi_{\rm o}^{4}(x), \,\,\,\,\,\,
k_{\perp} >  k_{0}. 
$$
Although  Eq. (\ref{spectrum0tt})
can be solved exactly for the transverse excitation spectrum \cite{book},
taking into account the coupling term  in the r.h.s. of  (\ref{spectrum0tt})
changes the values of excitation energies slightly, and  
the excitations 
can be approximately considered as of the pure excitonic 
($\hbar\omega_{1,k}= \hbar\omega_{{\rm ex},\,k_{\perp}}$) or the pure phonon 
($\hbar\omega_{2,k}= \hbar\omega_{{\rm ph},\,k_{\perp}}  $)  types.

It is  also useful to investigate {\it asymptotics} of the transverse inside excitations.
For definiteness sake, we investigate the left side  asymptotics of these  
excitations here, 
\begin{equation} 
{\rm u}_{j,s}({\bf x}) = \exp(\ell_{u}x/L_{0})\,{\rm e}^{ik_{\perp}y}\,u_{j,k} ,\,\,\,\,
\,\,\,{\rm v}_{j,s}({\bf x})  = \exp(\ell_{v}x/L_{0})\,{\rm e}^{ik_{\perp}y}\,v_{j,k},\,\,\,\,x<0,
\label{subst2}
\end{equation}
\begin{equation} 
C_{j,s}^{y}({\bf x})   = \exp(\ell_{c}x/L_{0})\, {\rm e}^{ik_{\perp}y} \,C_{j,k}^{y},
\,\,\,\,\,C_{j,s}^{x}({\bf x})  = \exp(\ell_{c}x/L_{0})\, {\rm e}^{ik_{\perp}y} \,C_{j,k}^{x}, 
\,\,\,\,x<0.
\label{subst22}
\end{equation}
Here \,$u_{j,k}$, $v_{j,k}$, $C_{j,k}^{y}$, and  $C_{j,k}^{x}$ 
are  smooth functions of ${\bf x}$ at \,
$\vert x \vert  > L_{0}$.   
Note that we introduced two components of ${ C}_{j,s}^{r} \sim {\rm e}^{ik_{\perp}y}$ 
to make Eqs. (\ref{eigfunction1})-(\ref{eigfunction4}) self-consistent. 
Let the equalities
\begin{equation}
\ell_{u}=\ell_{v}=1\,\,\,\,\,{\rm and}
\,\,\,\,\,1+\ell_{u}= \ell_{c}=2\
\label{combination}
\end{equation} 
be valid. 
Then the system of differential equations  (\ref{eigfunction1})-(\ref{eigfunction4}) can be reduced 
to a system of algebraic ones, which are analogous to Eqs. (\ref{EIGfunction01})-(\ref{EIG}).
Consequently, we can write out the equation for $\omega_{j,k}(x)$  valid at 
$\vert x \vert  > L_{0}$, 
$$  
\left(\,[\,\omega_{j,k} - iv\,(2/L_{0})\,]^{2}\,- \,c_{l}^{2} [k^{2}_{\perp}-(2/L_{0})^{2}]\,\right)
\times
$$
$$
\times \left[\, (\hbar\omega_{j,k} )^{2}-
\left(
\frac{ \hbar^{2} }{2m} [{ k}^{2}_{\perp}- \tilde{k}_{0}^{2} ]+
\tilde{F}(x) \right)\cdot 
\left( \frac{\hbar^{2} }{2m} [{ k}^{2}_{\perp} - \tilde{k}_{0}^{2} ] +
\tilde{F}(x)+2\nu_{0}(2\Phi_{\rm o}\exp(x/L_{0})\,)^{2}\right) \,\right]=
$$
\begin{equation}
=\left(\,
\frac{ \hbar^{2} }{2m} [{ k}^{2}_{\perp}-\tilde{k}_{0}^{2} ]+
\tilde{F}(x)\,\right)\, 2
\frac{\sigma_{0}}{M c_{l}^{2}}\,(\sigma_{0} a_{l}^{3})\,
 \bigl(2\Phi_{\rm o}\exp(x/L_{0} )\,\bigr)^{2} \,c_{l}^{2} [\,k^{2}_{\perp} - (2/L_{0})^{2}], 
\label{spectrum1t}
\end{equation}
where    \,$\tilde{k}_{0} =\sqrt{2}/L_{0}\simeq k_{0}$, \,$k_{\perp} >\tilde{k}_{0}$, and 
$$
\tilde{F}(x)=\vert \widetilde{\nu_{0}}\,(v)\vert
\left(\Phi_{\rm o}^{2} - \bigl( 2\Phi_{\rm o}\exp(x/L_{0})\, \bigr)^{2}\,\right)
\rightarrow   2\vert \tilde{\mu} \vert  - \epsilon 
 \,\,\,\, {\rm at}\,\,\,\, \vert x \vert \gg L_{0}.
$$ 
(We 
neglected     
the terms, such as $\widetilde{\nu_{1}}\,\phi_{\rm o}^{4}( x)\,{\rm u}_{j,s}({\bf x})$ 
and \,$\nu_{1}\phi_{\rm o}^{4}( x)\,{\rm v}_{j,s}({\bf x}) \sim \exp\bigl(\,(4+\ell_{u} ) x/L_{0}\,\bigr)$
in Eqs. (\ref{eigfunction1})-(\ref{eigfunction2}), and  the terms $\propto \kappa_{3}$
in Eq. (\ref{eigfunction4}) as well.)

Obviously, the structure of Eqs. (\ref{spectrum0tt}) and (\ref{spectrum1t}) is the same.
As the coupling between exciton and phonon branches
is weak for the transverse inside-excitations  (see the r.h. sides of 
Eqs. (\ref{spectrum0tt}) and  (\ref{spectrum1t})\,)
and the effect of the finite width $L_{0}$ can be taken into account as   
the spatial dependence of the important parameters in $\hbar\omega_{\rm ex}$,    
we use the following formula to estimate the low-energy excitation spectrum:
$$
(\hbar\omega_{{\rm ex},\,k_{\perp} })^{2}\simeq 
\left( \frac{\hbar^{2}}{2m}(k^{2}_{\perp} 
- k_{0}^{2})+F(x) +\epsilon_{+} \right)\, 
\left( \frac{\hbar^{2}}{2m}(k^{2}_{\perp} 
- k_{0}^{2}) + F(x) + 2\nu_{0}\phi_{\rm o}^{2}(x) 
+\epsilon_{+}'\right) \,\sim
$$
\begin{equation}
\sim \, \frac{\hbar^{2}}{2m}(k^{2}_{\perp} 
- k_{0}^{2})\,2\nu_{0}\Phi_{\rm o}^{2} 
+ 2\nu_{0}\Phi_{\rm o}^{2}\,\epsilon_{+} \,\,\,\,\, {\rm at}\,\,\,\,\, k_{\perp} \rightarrow k_{0}. 
\label{exciton11m}
\end{equation}
Here we take the inside-condensate-asymptotics of $F(x)$, $\phi_{\rm o}^{2}(x)$,
and $\epsilon_{+}  \approx \widetilde{\nu_{1}}(v)\, \phi_{\rm o}^{4}(x)$
to estimate $\hbar \omega_{\rm ex}$. 
Note that, for the inside-condensate
excitations,
 the the low-energy limit means\,  
$$
(\hbar^{2}/ 2m)(k^{2}_{\perp} 
- k_{0}^{2})  \simeq   (1 \sim 10)\,\vert \tilde{\mu} \vert. 
$$ 
Then, in the co-moving frame,  the low-energy excitation spectrum \,
$\hbar \omega_{{\rm ex}\,k_{\perp} }$ \,may develop a gap of the order of \,$\vert \tilde{\mu} \vert$,
(see  Eq. (\ref{OUTside}) for comparison).
Thus, inside the condensate,
we obtain a strong deviation 
of the collective excitation spectrum from both the simple excitonic 
one, \,$\vert\tilde{\mu}\vert +(\hbar^{2}/2m )k_{\perp}^{2} $, 
and the Bogoliubov-Landau spectrum \,$\propto \vert k_{\perp}\vert $.

\subsection{Longitudinal inside-excitations}
      
The case of  
the longitudinal excitations,  $k_{r}=k_{x}\,\delta_{1,r}$, 
$C_{j,{\bf k}}^{r}= C_{j, k}^{x}\,\delta_{1,r}$, is more difficult to analyze 
because
the mode interaction is 
non-negligible in the low-energy limit. 
(On Fig. 1., a longitudinal inside-excitation is shown with the two possible directions of the wave vector
${\bf k} \parallel Ox$.) 
Recall that the ``bare'' phonon modes, which can be written  in the laboratory frame as  
$$
u_{x}(x,t) \simeq q_{\rm o}(x-vt) + C_{k}^{x}(x-vt)\,{\rm exp}(ik_{x}x-i\omega_{\rm ph}t) + c.c.
$$
with \,$\omega_{\rm ph}=c_{l} \vert k_{x} \vert$ and 
$C_{k}^{x}(x) \sim \phi^{2}_{\rm o}(x)$, 
will be considered in the co-moving frame, $x-vt  \rightarrow  x$.
Then, within the inside-condensate approximation,  the following equation stands for the excitation 
spectrum:
$$  
\left(\,(\omega_{j,k} + vk_{x})^{2} \,- \,c_{l}^{2} k_{x}^{2}\,\right)\times
$$
$$
\times \left[\, (\hbar\omega_{j,k})^{2}-
\left(
\frac{ \hbar^{2} }{2m} [k_{x}^{2}-{ k}_{0}^{2} ]+
F(x)+ \epsilon_{+}\right)\cdot 
\left( \frac{\hbar^{2} }{2m} [ k_{x}^{2}-{ k}_{0}^{2} ] +
F(x)+2\nu_{0}\phi^{2}_{\rm o}(x) + \epsilon_{+}'\right) \,\right]=
$$
\begin{equation}
=\left(\,
\frac{ \hbar^{2} }{2m} [ k_{x}^{2}-k_{0}^{2} ]+
F(x)+ \epsilon_{+}\right)\,2 \frac{\sigma_{0}}{M c_{l}^{2}}\,(\sigma_{0}
 a_{l}^{3})\,\Phi_{\rm o}^{2} \,(c_{l}^{2} k_{x}^{2}), \,\,\,\,\,\, \vert x \vert < L_{0}.
\label{spectrum0ll}
\end{equation}

It is important to note that, unlike the case of transverse excitations,  the values of 
$$
\Bigl( \hbar\omega_{{\rm ph},\, k_{x}}^{(0)} \Bigr)^{2}\simeq \hbar^{2}\,(c_{l}-v)^{2}
\bigl((3\sim 7)\,k_{0}\bigr)^{2} 
$$
and   
$$
\Bigl(\hbar \omega_{{\rm ex},\, k_{x}}^{(0)} \Bigr)^{2}
\simeq
\left( \frac{\hbar^{2}}{2m}(10\sim 40) 
k_{0}^{2} +\epsilon_{+} \right)\, 
\left( \frac{\hbar^{2}}{2m} 
(10\sim 40) k_{0}^{2} + 2\,\nu_{0}\,\phi_{\rm o}^{2}(x) 
+\epsilon_{+}'\right)
$$
are of the same order of value at $k_{x} \simeq  (3 \sim 8)\,k_{0}$, and 
the inequality
$
\bigl(\hbar \omega_{{\rm ex}, \,k_{x}}^{(0)} \bigr)^{2}>
\bigl( \hbar\omega_{{\rm ph}, k_{x}}^{(0)} \bigr)^{2}
$
is valid in the low-energy limit.
Moreover, the two cases, $k_{x}>0$ ($+$\,-\,case) and $k_{x}<0$ ($-$\,-\,case), 
are different as it can be seen from  
the l.h.s. of Eq. (\ref{spectrum0ll}).
In the low-energy limit,
we can write the approximate solution of  (\ref{spectrum0ll})
as follows
$$
\Bigl( \hbar \omega_{{\rm ex}, \,k_{x}}^{(\pm)}\Bigr)^{2} \approx
\left(
\frac{ \hbar^{2} }{2m} [k_{x}^{2}-{ k}_{0}^{2} ]+
F(x)+ \epsilon_{+}\right)\times
$$ 
$$
\times
\left( \frac{\hbar^{2} }{2m} [ k_{x}^{2}-{ k}_{0}^{2} ] +
F(x)+ 2\nu_{0}\,\phi^{2}_{\rm o}(x) \pm  
2 \,q_{\pm} \gamma(v)\,\frac{\sigma_{0}}{M c_{l}^{2}}\,(\sigma_{0}
 a_{l}^{3})\,\phi_{\rm o}^{2}(x) 
+  \epsilon_{+}'\right),
$$   
where $q_{+} \sim 1$ and $0<q_{-} < 1$.
Note that $\hbar \omega^{(+)}_{{\rm ex}, \,k_{x}} >\hbar \omega_{{\rm ex}, \,k_{x}}^{(0)}$,
whereas,
for the phonon-type branch, 
$\omega^{(+)}_{{\rm ph},\, k_{x}} < \,(c_{l}-v)\,k_{x}$.
For $k_{x}< 0$,  we have the following  inequaluty
for the excitonic branch,
$$
\left(
\frac{ \hbar^{2} }{2m} [k_{x}^{2}-{ k}_{0}^{2} ]+
F(x)+ \epsilon_{+}\right)\times
$$
\begin{equation}
\times \left( \frac{\hbar^{2} }{2m} [ k_{x}^{2}-{ k}_{0}^{2} ] +
F(x)+ 2\,\widetilde{\nu_{0}}(v)\,\phi^{2}_{\rm o}(x)  
+  \epsilon_{+}'\right) <
\Bigl(\hbar \omega^{(-)}_{{\rm ex}, \,k_{x}} \Bigr)^{2}< 
\Bigl(\hbar \omega_{{\rm ex}, \,k_{x}}^{(0)} \bigr)^{2},
\label{INequality1}
\end{equation}
where $2\,\widetilde{\nu_{0}}(v)\,\phi^{2}_{\rm o}(x) \simeq - 4\,\vert \tilde{\mu} \vert$ 
within the inside-approximation,   
and, for the phonon-type branch, we obtain  
$ \omega^{(-)}_{{\rm ph},\, k_{x}} > \,(c_{l}+ v)\,\vert k_{x} \vert$.

To derive the formulas for the amplitudes \,${\rm u}_{ k}({ x})$, \,${\rm v}_{ k}({ x})$, and 
\,${ C}_{ k}^{x}({ x})$ of the excitonic branch, we use the following approximations 
$$
L_{\pm}(-{ k}_{x}^{2})=L(-{ k}_{x}^{2})+ \frac{\rho^{-1}\sigma_{0}^{2}
\phi_{\rm o}^{2}(x)\, k_{x}^{2} }{ (\omega_{{\rm ex},\,k_{x}} \pm v\vert k_{x}\vert )^{2} - c_{l}^{2} k_{x}^{2} }
\approx L(-{ k}_{x}^{2})\,\pm \,q_{\pm} \gamma(v)\,\frac{\sigma_{0}}{M c_{l}^{2}}\,(\sigma_{0}
 a_{l}^{3})\,\phi_{\rm o}^{2}(x),  
$$
and $B = \nu_{0}\phi_{\rm o}^{2}({ x}) + 2\nu_{1}\phi_{\rm o}^{4}({ x})$ is  modified as  
$$
B_{\pm} \approx \nu_{0}\phi_{\rm o}^{2}({ x}) + 
2\nu_{1}\phi_{\rm o}^{4}({ x}) \pm q_{\pm} \gamma(v)\,\frac{\sigma_{0}}{M c_{l}^{2}}\,(\sigma_{0}
 a_{l}^{3})\,\phi_{\rm o}^{2}(x). 
$$
Then we can rewrite the formulas for the excitonic excitation spectrum as  follows
$$ 
\bigl( \hbar \omega_{{\rm ex}, \,k_{x}}^{( \pm )}  \bigr)^{2} \approx 
  L_{\pm }^{2}(-{ k}_{x}^{2})  -  B_{\pm}^{2}=
 \Bigl(L(-{ k}_{x}^{2}) -  B \Bigr)\Bigl( L_{\pm }(-{ k}_{x}^{2})+  B_{\pm} \Bigr).
$$
Recall that the orthogonality relation  (\ref{impORT22}) can be used to normalize the amplitudes.  
Within the inside-approximation,  Eq. (\ref{impORT22})  can be rewritten 
as follows ($\delta _{ss}=1  \rightarrow \delta _{kk}=1$) 
\begin{equation}
\int\!d{\bf x}\,\bigl( \vert {\rm u}_{k}({ x})\vert ^{2} - 
\vert {\rm v}_{k }({ x}) \vert ^{2}\,\bigr) \,+\,
(1/\hbar)\!\int\!d{\bf x}\, 
2\rho\, (\omega_{{\rm ex}, \,k_{x}}  + vk_{x}) 
\vert C_{k}^{x} ( x)    \vert ^{2}  = 1, 
\label{ortUVC1}
\end{equation}
and  we have for the excitonic amplitudes
\begin{equation}
\Bigl\vert
{\rm u}_{ k}^{(\pm)}(x)
\Bigr\vert^{2}
\approx  \left(\frac{ \Upsilon_{\pm} }{V_{\rm eff}}\right)\frac{ L_{\pm } (-k_{x}^{2}) 
+\hbar\omega_{{\rm ex}, \,k_{x}}^{( \pm )}    }{2\,\hbar\omega_{{\rm ex}, \,k_{x}}^{( \pm )} },
\,\,\,\,\,
\Bigl\vert
{\rm v}_{k }^{(\pm)}  (x)
\Bigr\vert^{2}
\approx \left(\frac{\Upsilon_{\pm}}{V_{\rm eff}}\right)
\frac{ L _{\pm}(-k_{x}^{2}) -\hbar\omega_ {{\rm ex}, \,k_{x}}^{( \pm )} }{
2\,\hbar\omega_{{\rm ex}, \,k_{x}}^{( \pm )} },
\label{uvAmplitude}
\end{equation}
$$
{\rm u}_{ k}^{(\pm)\,*}{\rm v}_{k }^{(\pm)}  (x)
\approx
-\left(\frac{\Upsilon_{\pm}}{V_{\rm eff}}\right)\frac{B_{\pm} }{2\,\hbar\omega_{{\rm ex}, \,k_{x}}^{( \pm )}    }.
$$
Here  the effective condensate 
volume $V_{\rm eff} \simeq  2SL_{0} $ is used to normalize 
the u- and v-wave functions of the inside 
excitations, and 
\,$\int\!\!d{\bf r}\,(\vert {\rm u}_{k}\vert ^{2} - \vert {\rm v}_{k}\vert ^{2})=\Upsilon_{\pm } < 1$. 

Subsiquently, we get for 
\begin{equation}
C_{k}^{x}(x) =-
\frac{\rho^{-1}\sigma_{0}\,\phi_{\rm o}(x)\,i k_{x}\,\bigl({\rm u}_{k}(x) +{\rm v}_{k}(x)\,\bigr)}
{  (\omega_{{\rm ex}, \,k_{x}}+vk_{x})^{2} - c_{l}^{2}{ k}^{2} }
\end{equation}
the following approximate formulas 
$$
C_{k}^{x\,(\pm)}(x) \approx \mp \,q_{\pm}\, \gamma(v)\,\frac{\sigma_{0}}{M c_{l}^{2}}\,
\sqrt{ a_{l}^{3}\,\phi_{\rm o}^{2}(x) } \,\frac{i}{k_{x}}\,\sqrt{ a_{l}^{3}}\,
\bigl({\rm u}_{k}^{(\pm)}(x) +{\rm v}_{k}^{(\pm)}(x)\,\bigr). 
$$
To estimate the characteristic value   of \,$C_{k}^{x\,(\pm)}(x)$, we use  
${\rm u}_{k}(x) \simeq  {\rm v}_{k}(x) \sim \sqrt{ \Upsilon_{\pm} /V_{\rm eff} }$
and obtain 
$$
\vert C_{k}^{x\,(\pm)} \vert \simeq  q_{\pm}\, \gamma(v)\,\frac{\sigma_{0}}{M c_{l}^{2}}\,
\sqrt{ a_{l}^{3}\,\Phi_{\rm o}^{2} }\,\sqrt{\frac{ \Upsilon_{\pm}\,a_{l}\,L_{0} }{2S   } }\,
 \frac{ a_{l}}{(3\sim 7)} <\!\!<\!\!<  a_{l}.
$$

The parameters $\Upsilon_{\pm}$ characterize the relative weight of excitonic degrees of freedom 
in the considered branch of excitations.
As 
the parameter \,$\hbar vk_{x}/\hbar\omega_{{\rm ex}, \,k_{x}}^{(+)} < 1$  
at \,$k_{x} \simeq (4 \sim 8)\,k_{0}$, 
the parameter \,$\Upsilon_{+}(k_{x}) $ \,can be estimated as $0.5 \sim 0.7$. 
For $k_{x}<0$, we obtain the following equation from (\ref{ortUVC1}),
$$
\Upsilon_{-}\, \left(1 + \frac{ \bigl( a_{l}^{3}\,\Phi_{\rm o}^{2}\bigr) }{ m /M }
\,q_{-}^{2}\,\gamma^{2}(v)\, \Bigl(\frac{\sigma_{0}}{M c_{l}^{2}}\Bigr)^{2}
\left(1 -  \frac{ \hbar v\vert k_{x}\vert  }{ \hbar\omega_{{\rm ex}, \,k_{x}}^{(-)} } \right)
 \frac{ L(-{ k}_{x}^{2}) -  B}{ \hbar^{2}k_{x}^{2}/2m } 
\,\right)
\approx 1.
$$
Within the stability area (see the next subsection for an extended discussion), 
we estimate
the ratio $\hbar v\vert k_{x}\vert /\hbar\omega_{{\rm ex}, \,k_{x}}^{(-)} \simeq 1/2 \sim 1/3 $ 
at \,$\vert k_{x}  \vert \simeq (4 \sim 8)\,k_{0}$. Then,
\, $\Upsilon_{-}(k_{x}) >0$ and  \,$\Upsilon_{-} \simeq  0.6 \sim 0.8 $.

To go beyond the  inside-approximation, 
the effect of inhomogeneous  behavior of the longitudinal excitations
can be considered.
We use the following {\it ansatz} for the left side asymptotics (see Fig. 2)   
 \begin{equation} 
{\rm u}_{j,s}({\bf x}) = \exp(\ell_{u}x/L_{0})\,{\rm e}^{ik_{x}x}\,u_{k} ,\,\,\,\,
\,\,\,{\rm v}_{j,s}({\bf x})  = \exp(\ell_{v}x/L_{0})\,{\rm e}^{ik_{x}x}\,v_{k},
\label{subst2ll}
\end{equation}
\begin{equation} 
C_{j,s}^{x}({\bf x})  = \exp(\ell_{c}x/L_{0})\, {\rm e}^{ik_{x}x} \,C_{k}^{x}, \,\,\,\,\,\,x<0.
\label{subst22ll}
\end{equation}
where \,$ \ell_{u}=\ell_{v}=1$ 
\,and\, $ \ell_{c} =2$.
Then, like the case of transversal excitations (see Eq. (\ref{spectrum1t})), 
we can write  the equation for $\omega_{j,\,k_{x}}(x)$  valid at 
$\vert x \vert  > L_{0}$, 
$$  
\left(\,(\,\omega_{j,\,k_{x}} + v\,\tilde{k}_{x})^{2}\,- \,c_{l}^{2} \tilde{k}^{2}_{x}\,\right)
\times
$$
$$
\times \left[\, (\hbar\omega_{j,\,k_{x}} )^{2}-
\left(
\frac{ \hbar^{2} }{2m} [\bar{ k}^{2}_{x} - {k}_{0}^{2} ]
+ \tilde{F}(x) \right) 
\left( \frac{\hbar^{2} }{2m} [\bar{ k}^{2}_{x} - {k}_{0}^{2} ] +
\tilde{F}(x)+2\nu_{0}\bigl(2\Phi_{\rm o}\exp(x/L_{0})\,\bigr)^{2}\right) \,\right]=
$$
\begin{equation}
=\left(\,
\frac{ \hbar^{2} }{2m} [\bar{ k}^{2}_{x}- {k}_{0}^{2} ]+
\tilde{F}(x)\,\right)\, 2
\frac{\sigma_{0}}{M c_{l}^{2}}\,(\sigma_{0} a_{l}^{3})\,
 \bigl(2\Phi_{\rm o}\exp(x/L_{0} )\,\bigr)^{2} \,c_{l}^{2}\tilde{k}^{2}_{x}, 
\label{spectrum1ll}
\end{equation}
where  
$k_{x} \rightarrow \tilde{k}_{x}=k_{x}- i(2/L_{0})$ in the phonon parts of this equation and 
$k_{x} \rightarrow 
\bar{k}_{x} =k_{x}-i(1/L_{0})$  in the exciton parts  of it
($x<0$).
It is easy to see that Eqs. (\ref{spectrum0ll}) and (\ref{spectrum1ll})
are in the continuity correspondence,  i.e. they describe the same object.
For example, the (left side) asymptotic behavior of 
$\hbar\omega_{{\rm ex}, \,k_{x}}^{(\pm)}(x)$ can be obtained 
from  the  inside-condensate formulas
by the substitute  
$$ k_{x} \rightarrow  \bar{k}_{x},\,\,\,\,\,
F(x)\rightarrow \tilde{F}(x) \rightarrow 2\vert \tilde{\mu} \vert, 
$$ 
and 
$$
\phi_{\rm o}^{2}(x) \rightarrow (2\Phi_{\rm o}\exp(x/L_{0})\,)^{2}   \rightarrow 0.
$$
As a result, we obtain
$\hbar \omega_{{\rm ex}, \,k_{x}  }  \simeq \vert \tilde{\mu} \vert + \hbar^{2}\bar{k}^{2}_{x}/2m$
that corresponds to the outside-excitation spectrum, Eq. (\ref{gap1}).

The  excitonic input into  $\langle\delta\hat{\Psi}^{\dag} \delta\hat{\Psi} ({\bf x})\rangle_{T=0}$, 
the quantum depletion
of the moving condensate,  
can be calculated by  $\sum_{1,s} \vert {\rm v}_{1,s} (x) \vert^{2} $ \cite{Pit}.
To estimate this value,  one can approximate
$\vert {\rm v}_{k }(x) \vert^{2}$ as follows
$$
\Bigl\vert
{\rm v}_{k_{x} }^{(\pm)}  (x)
\Bigr\vert^{2}
\simeq  \left(\frac{\Upsilon_{\pm}}{V_{\rm eff}}\right)
\frac{B_{\pm}^{2} }{ 4L _{\pm}^{2}(-k_{x}^{2})}. 
$$
However, the summation $\sum_{1,s}$\, implies \,$\int \!dk_{x}\,d^{2}k_{\perp}/(2\pi)^{3}$ within the semiclassical
approximation. 
Assuming that such an integration makes the difference among  ${\rm v}_{k_{x} }^{(-)}$,\, 
${\rm v}_{k_{x} }^{(+)}$ \,and\, 
${\rm v}_{k_{\perp} }$ not essentially  important, 
we  use the following estimate  for  $\vert {\rm v}_{\bf k }(x) \vert^{2}$,
$$
\Bigl\vert
{\rm v}_{ \bf k } (x)
\Bigr\vert^{2}
\simeq  \left(\frac{1 }{V_{\rm eff}}\right)
\frac{B^{2} }{ 4L^{2} (-{\bf k}^{2})}. 
$$
Then, the integration $\int \!dk_{x}\,d^{2}k_{\perp}/(2\pi)^{3}$ can be reduced to 
$\int_{k_{0}} \!k^{2} dk/2 \pi^{2}$, and  the main input ($\sim  \phi_{\rm o}^{2}(x)$)
can be estimated  from the following formula
$$
\langle \delta\hat{\Psi}^{\dag} \delta\hat{\Psi}({\bf x})\rangle_{T=0} \,
\simeq \frac{1}{8 \pi^{2} \, L_{0}^{3} }\,\frac{\nu_{0}\phi_{\rm o}^{2}(x)}{ 2 \vert \tilde{\mu}\vert }
 +  \frac{\epsilon }{8 \pi^{2} \, L_{0}^{3} }\,\frac{\bigl(\nu_{0}\phi_{\rm o}^{2}(x)\,\bigr)^{2}}{ 2 \vert \tilde{\mu}\vert^{2} }
\sim 
\frac{1}{8  \pi^{2}\, L_{0}^{3} }\,\frac{\nu_{0}\Phi_{\rm o}^{2} \,{\rm cosh}^{-2}(x/L_{0})}{ \vert \widetilde{\nu_{0}}\vert \,\Phi_{\rm o}^{2} }.
$$
Using this density, we speculate that       
 the {\it localized} depletion of condensate -- 
i.e. the number of particles that are out of the condensate but move with it coherently --
seems to be a small value.  We obtain the following estimate  ($\nu_{0} = \varepsilon_{0}\,\tilde{a}_{B}^{3}$ \,and\,
$\vert \widetilde{\nu_{0}}(v)\vert  =  \vert \tilde{\varepsilon}_{0}(v)\vert\,\tilde{a}_{B}^{3}$):
$$
\delta N_{0} = \int\!d{\bf x}\,\langle \delta \hat{\Psi}^{\dag} \delta \hat{\Psi} ({\bf x})\rangle_{T=0}\, \simeq  \,
 \frac{1}{16\,\pi^{2} }\,\frac{\varepsilon_{0}}{\vert \tilde{\mu}\vert }\,
 \frac{\tilde{a}_{B}^{3} }{L_{0}^{3}}\, N_{\rm o}, 
$$
where  the factor before $N_{\rm o}$  can be estimated as 
$$
\frac{\varepsilon_{0}}{\vert \tilde{\mu}\vert (N_{\rm o},\,v) }\,
 \frac{\tilde{a}_{B}^{3} }{L_{0}^{3}(N_{\rm o},\,v) } \simeq 
\frac{\varepsilon_{0} \,\vert\tilde{\varepsilon}_{0}(v) \vert }{2\,x\,({\rm Ry}^{*})^{2} }
\, \frac{1}{\bar{n}_{\rm o}} \sim (10^{-1} \sim 10^{-2})\,\bar{n}_{\rm o}^{-1}.
$$
(Here we used  Eqs. (\ref{need1}),(\ref{need2}) within the approximation \,$\bar{n}_{\rm o} \gg 10$.)
Note that 
there is no small ${\bf k}$ input to the estimate of $\sum_{1,s}$ because, 
first,   
such excitations belong to the outside-excitation branch in our model, and, second, we use
the approximation \,${\rm v}_{\bf k}(x) \approx 0$ for them.


\subsection{Stability of the Moving Condensate}

To investigate the stability of the moving condensate in 
relation to the creation of inside-excitations, we calculate the energy 
of the condensate with the {\it one} inside excitation,
$\langle   \alpha^{\dag}_{1,s}  \alpha_{1,s} \rangle =1$, 
described by 
the  following set: $k$, $\omega_{k}$,  and ${\rm u}_{k}$, ${\rm v}_{k}$, $C_{k}$. 
In this study,  we analyze the stability inside  the excitonic sector of our model.
 
Although the  excitations
were defined  in the co-moving frame, calculations should be done in the laboratory frame. 
Returning to the lab frame, we represent
the exciton and phonon field functions as follows:   
\begin{equation}
\phi_{\rm o}(x-vt,\,t) \rightarrow \phi_{\rm o}(x-vt,\,t) + \exp\bigl(-i(\tilde{E}_{g} + mv^{2}/2 
-\vert \tilde{ \mu }\vert) t\,\bigr)\exp\bigl(i(\varphi + mvx)\,\bigr)\,\delta\tilde{\Psi}({\bf x}, t),
\label{urfin}
\end{equation}
where
$$
\delta\tilde{\Psi}({\bf x}, t)=
{\rm u}_{k}(x-vt)\,{\rm e}^{ i(\varphi_{0} +{\bf k}{\bf x})} \,{\rm e}^{-i(\omega_{k}+k_{x}v)t } +
{\rm v}_{k}(x-vt)\,{\rm e}^{-i(\varphi_{0} + {\bf k}{\bf x})}\,{\rm e}^{i(\omega_{k}+k_{x}v)t},  
$$
and 
\begin{equation}
u_{\rm o}(x-vt) \rightarrow u_{\rm o}(x-vt) + 
C_{k}(x-vt)\,\exp\bigl(i(\varphi_{0}+ {\bf k}{\bf x})\bigr)\,\exp\bigl(-i(\omega_{k}+k_{x}v)\,t \bigr)+ {\rm c.c.},
\label{urfin1}
\end{equation}
see Eq. (\ref{OUTside}) for comparison.
In this analysis, the inside excitations 
are not considered as fluctuations, and 
the (average) number of particles in the condensate  and its energy  are changed as 
$N_{\rm o} - \int \!d{\bf x}\, \delta\psi^{\dag}\delta\psi$ and 
\,$E_{\rm o}(N_{\rm o}) - (\partial_{N}E_{\rm o})\int\! d{\bf x}\,\delta\psi^{\dag}\delta\psi$,
respectively.  However, these changes are not important if the number of excitation in a system 
is less than $\sqrt{\delta N^{2}} \simeq \sqrt{N_{\rm o}}$. 
They could be important  in the case
of instability of the moving condensate.

The zeroth component 
of the energy-momentum tensor can be represented in the form   
$$
{\cal T}_{0}^{0}={\cal T}_{0}^{0}(\phi_{\rm o}, \,u_{\rm o})+ 
{\cal T}_{0}^{0\,(2)}(\delta\Psi^{\dag}, \delta\Psi, \delta u,\,\partial_{t}\delta u\,
\vert\,\phi_{\rm o}, \,u_{\rm o}),
$$
where the first part corresponds to the condensate energy $E_{\rm o}$ and the second
part gives the energy of inside-excitations, $E_{\rm in}$. 
After substitution of (\ref{urfin}),(\ref{urfin1}) 
into $E_{\rm in}=\int \!d{\bf x}\,{\cal{T}}_{0}^{0\,(2)}$,  
we have for the total energy  
$$
E_{\rm o}+E_{\rm in-ex}\, \approx 
$$
$$
\approx\,
E_{\rm o}(N_{\rm o}) 
+  \delta E_{\rm o}(N_{\rm o})
+
\int\!\! d{\bf x}\,\hbar(\omega_{k}(x) + k_{x}v)\,\left\{\vert {\rm u}_{k} \vert^{2} -
\vert {\rm v}_{k}\vert^{2} +
(2/\hbar) \rho(\omega_{k}(x) + vk_{x})
\vert C_{k} \vert ^{2} \right\}\, =
$$
\begin{equation}
=\, E_{\rm o}(N_{\rm o}) +  \delta E_{\rm o}(N_{\rm o})
+ \hbar\,\langle\omega_{k} + k_{x}v\rangle,
\label{stable}
\end{equation}
where 
\begin{equation}
\delta E_{\rm o}(N_{\rm o}) = (2\vert\tilde{\mu}\vert + \nu_{0}\Phi_{\rm o}^{2}) 
\int\!\! d{\bf x}\,\delta\tilde{\psi}^{\dag}
\delta\tilde{\psi} - 
 \frac{M(c_{l}^{2}+v^{2})}{2}\,3\vartheta(N_{\rm o},\,v) \int \!\!d{\bf x}\,\delta\tilde{\psi}^{\dag}\delta\tilde{\psi},
\label{stableornot}
\end{equation}
$$
\delta\tilde{\psi}^{\dag}\delta\tilde{\psi}(x) \rightarrow 
\vert {\rm u}_{k}\vert^{2}+\vert {\rm v}_{k} \vert^{2},
$$
see Eqs. (\ref{cond_and_exc1}),(\ref{Cond_And_Exc1}) for comparison.

In this study, we discuss qualitatively 
the stability of the condensate in relation to the backward emission of inside-excitations,
(i.e. $k_{x} < 0$ in the co-moving frame). 
To begin with, we consider the standard criterion, 
\begin{equation}
\hbar (\omega^{(-)}_{ {\rm  ex},\,k} -   
\vert  k_{x} \vert  v) >0 \,\,\,\,\,{\rm at }\,\,\,\,\,\,\vert  k_{x} \vert \simeq  z\,L_{0}^{-1},
\label{STable}
\end{equation}   
where  $z \simeq  3 \sim 10$ corresponds to the low-lying excitations.
The value of $\omega^{(-)}_{ {\rm  ex},\,k}(x)$ is taken within the 
inside-approximation, see Eq. (\ref{INequality1}),  so that 
$\hbar \omega^{(-)}_{ {\rm  ex},\,k} \simeq f(z)\,\vert \tilde{\mu}\vert$ 
and \,$z / f(z) \simeq 0.1 \sim 0.3$.
Then,
it is easy to conclude that the following inequality  
\begin{equation}
\frac { \hbar \vert  k_{x} \vert  v }{ \hbar \omega^{(-)}_{ {\rm  ex},\,k} }
 \simeq  \frac{z}{f(z)}\,\frac{v}{c_{l}}\,\sqrt{ \frac { (4m/M)\, Mc_{l}^{2}/2 }
{ \vert \tilde{\mu}\vert (N_{\rm o},\,v)   } }<1
\label{Stable1}
\end{equation}
is valid in the low-energy limit 
if the effective chemical potential (\ref{need2})
is large enough, see Eq. (\ref{Stable}) for comparison. 

More precisely,  the ballistic velocity $v$ and 
the number of particles in the condensate, $N_{\rm o}$,  
have to be large enough, f.ex.,     
$\vert \tilde{ \varepsilon}_{\rm o}(v) \vert \simeq (10^{-1} \sim 1)\,{\rm  Ry}^{*}$ and 
$\bar{n}_{\rm o} \simeq 10$, in order to the inequality
\begin{equation} 
\frac{ (4m/M)\, Mc_{l}^{2}/2 }{ \vert \tilde{\mu} \vert (N_{\rm o},\,v) } 
\simeq \frac{ (4m/M)\, Mc_{l}^{2}/2 }{
\vert \tilde{ \varepsilon}_{\rm o}(v) \vert^{2}\,/\,4\bar{n}_{\rm o}^{2}\,x\,{\rm  Ry}^{*} }
< 10 \sim 20
\label{real_param}
\end{equation}
can be satisfied.
Thus,  for  
\begin{equation}
\vert \tilde{\mu}\vert (N_{\rm o},\,v)  >  
\mu_{\rm cr} \simeq  10^{-1}\,(4m/M)\, Mc_{l}^{2}/2, 
\label{cr_mu}
\end{equation}
where $\mu_{\rm cr} \sim 10^{-4}$\,eV, 
one can expect conditions  (\ref{Stable1}),(\ref{real_param}) to be  valid.  

Despite the condensate can be formed near 
$\gamma (v) \approx \gamma_{\rm o}$ in theory, f.ex., 
with $\vert \widetilde{\nu_{0}}(v) \vert  \simeq  0.1\,\nu_{0}$ 
and $\bar{n}_{\rm o} \gg 10$,
such a ballistic state seems to be unstable against the creation of inside-excitations.
Note that the critical (Landau) velocity, $v_{\rm cr}$, can be found as a solution of Eq.  (\ref{cr_mu}) and 
$v_{\rm o} < v_{\rm cr}(N_{\rm o}) < c_{l}$. In fact,   
the parameter $\vert \tilde{\mu} \vert / \mu_{\rm cr}$ controls the stability/instability of the condensate, 
see Eqs. (\ref{Stable}),(\ref{Stable1}). 

Analyzing (\ref{STable}), we did not take into account  Eq. (\ref{stableornot}). 
However, if the instability regime takes place, more than $\sqrt{N_{\rm o}}$ inside-excitations can 
appear. As  
the changes in $u_{\rm o}(x-vt)$ 
because of  $N_{\rm o} \rightarrow N_{\rm o} -\delta N $
are nonlocal (in spite of creation of the localized excitations, see Figs. 1 and 3),  a  free acoustic phonon 
can appear in the system lattice\,$+$\,excitons {\it together} with appearance of the  localized excitation 
$\hbar \omega^{(-)}_{ {\rm  ex},\,k}(x)$.
Like the case of outside-excitations, 
we assume that  
$(3/2)\,M(c_{l}^{2}+v^{2})\,\vartheta(N_{\rm o},\,v) \sim  \hbar c_{l} k_{\rm ph}$, see  
Eq.  (\ref{stableornot}). Then, only the term  $ \propto  2\vert\tilde{\mu}\vert + \nu_{0}\Phi_{\rm o}^{2}$
is important. In fact, this term leads to some renormalization of the values of the critical parameters, 
$\mu_{\rm cr}$ and $v_{\rm cr}$. 

\newpage
\begin{figure}
\begin{center}
\leavevmode
\epsfxsize = 410pt
\epsfysize = 300pt
\epsfbox{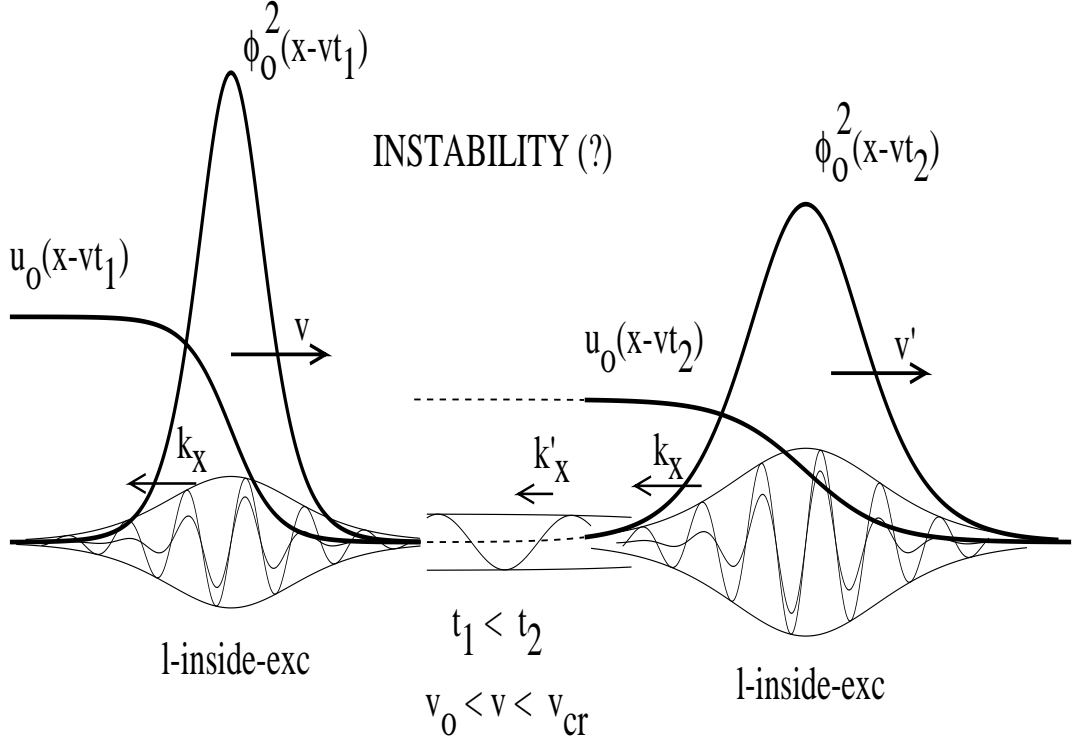}
\end{center}
\caption{
The ballistic condensate,
$\phi_{\rm o}(x-vt)\cdot u_{\rm o}(x-vt)\delta_{1j}$, can be unstable 
in relation to emission of the inside-excitations
if the effective chemical potential $\vert \tilde{\mu }\vert (N_{\rm o},\,v)\,<\,\mu_{cr}$.
In terms of Landau critical velocity, this means    
$v_{\rm o}< v < v_{\rm cr}(N_{\rm o})$.
If such an instability takes place, the emission of inside-excitations can be accompanied by 
the emission of outside-excitations of the condensate.  The longitudinal inside-excitations
are labeled by the wave vector $k_{x} < 0$ on this figure, whereas the outside-excitation
is labeled by  the wave vector $k'_{x} < 0$.
}
\label{l-inside1}
\end{figure}

\section{Interference Between Two Moving Packets} 

In this section, we address the problem of interaction between two moving 
condensates. This problem is
essentially nonstationary, 
especially if the initial ballistic velocities of packets are different. 
Within the  quasi-1D conserving model, the following equations govern the dynamics
of the two input packets  (we choose the reference frame moving with the slow packet): 
\begin{equation}
\Bigl(\,i\hbar\partial_{t} + \frac{ m(v')^{2}}{2}\,\Bigr)\psi_{0}(x,t)= \Bigl(-\frac{\hbar^{2}}{2m }\partial^{2}_{x}+
\nu_{0}\vert\psi_{0}\vert^{2} + \nu_{1}\vert\psi_{0}\vert ^{4}\Bigr)\,\psi_{0}(x,t) + 
\sigma_{0}\,\partial_{x} u_{0}(x,t)\,\psi_{0}(x,t),
\label{timedep1D}
\end{equation}
\begin{equation}
\bigl(\,(\partial_{t}- v'\,\partial_{x})^{2} -  c_{l}^{2} \partial_{x}^{2} 
\bigr)\,u_{0}( x,t)-c_{l}^{2}\,2\kappa_{3}\,\partial_{x}^{2}u_{0}( x,t)\,\partial_{x} u_{0}( x,t)
=
\rho^{-1}\sigma_{0}\,\partial_{x}\vert\psi_{0}\vert^{2}( x,t).
\label{timedep1DD}
\end{equation}
Then, the initial conditions  can be written in the explicit 1D form by using  the
exact solution  of the model (\ref{1Deq}),(\ref{11Deq}). Note that the amplitudes 
of the stationary ballistic state, $\phi_{\rm o}(x-vt) \cdot \partial_{x} u_{\rm o}(x-vt)$,
were defined from the normalization condition  and depend on
the values of $v$ and $N_{\rm o}$. Hence,
the amplitudes of the ``input'' condensates for 
Eqs. (\ref{timedep1D}),(\ref{timedep1DD}) may not 
have the same values.

\begin{figure}
\begin{center}
\leavevmode
\epsfxsize = 400pt
\epsfysize = 300pt
\epsfbox{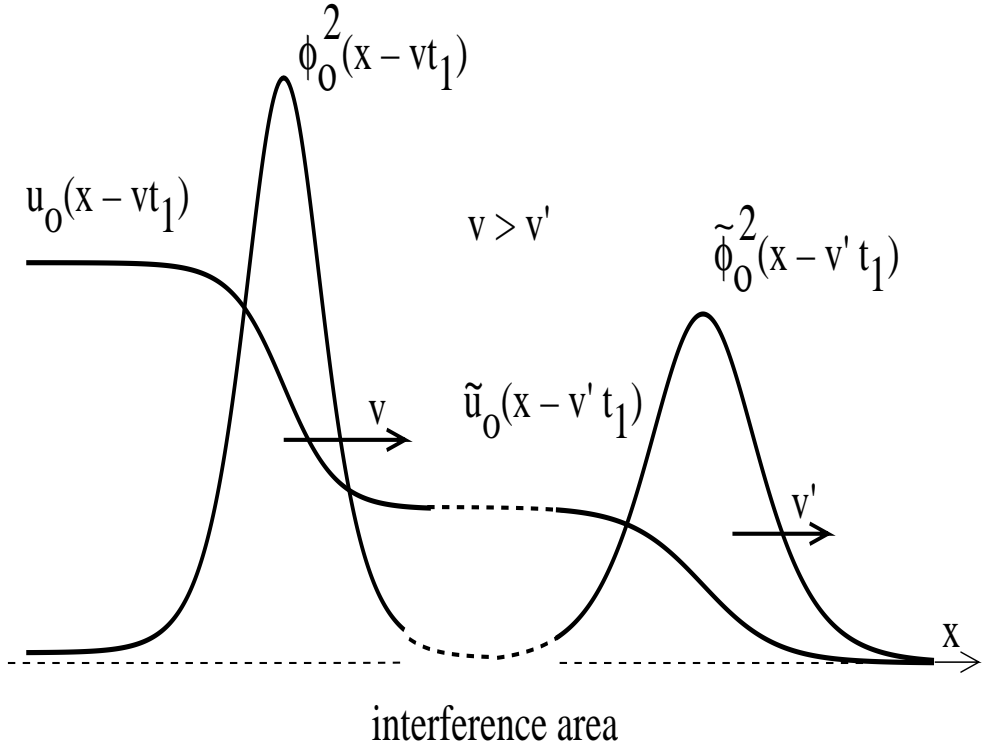}
\end{center}
\caption{
Two ballistic condensates move with different velocities, 
$v-v' \simeq (0.1\sim 0.3)\,c_{l}$,  \,and \,$t=t_{1}$  before 
the ``collision'', or, the strong interaction process.
 If one can prescribe the coherent phase to 
each of the  participating condensates, e.g., $\varphi_{\rm c}({\bf x}) \approx \varphi + (mv/\hbar)\,x$,  
the interference area appears between them.
(The interference area is marked by bold dashed lines on this figure.) 
As $v \ne v'$, the fringes are non-stationary, and  the outside-excitations 
can actually be excited in this area. 
}
\label{l-inside11}
\end{figure}

In this study, we approach the problem of strong interaction
between the condensates.  Therefore, we choose 
the nonsymmetric initial 
conditions, i.e. 
the amplitude and the velocity of the ``input'' 
packets are different, for example, $v>v'$ and 
${\bf v}\|{\bf v}'$, see Fig. \ref{l-inside11}.
Here, we reply on the experimental observation \cite{Fortin}
that, at $v>v_{\rm o}$, the 
ballistic velocity of the condensate depends on the  power of a laser beam
irradiating  the crystal. 
 If the exciton concentration in the first packet, $n_{{\rm o}\,1}$, is close to the 
value of  the Bose condensation threshold,
the exciton concentration in
the second packet, $n_{{\rm o}\,2} > n_{{\rm o}\,1}$,
the velocity difference between condensates can 
reach  
$(0.1 \sim 0.3)\,c_{l}$.  
Then, in the reference frame moving with the first (slow) packet, the initial
conditions can be taken as the following: 
 $$
 \psi_{0}({ x},t=0)\cdot { u}_{0}({ x},t=0)= 
  \phi_{\rm o}( x;N_{{\rm o}\,1})\cdot { q}_{\rm o}({ x};N_{{\rm o}\,1})\,+\,
  $$
  \begin{equation}
 +\exp\bigl( i(\delta\varphi + m\,\delta  v \, x)\bigr)\,
 \phi_{\rm o}({ x}+{ x}_{0};N_{{\rm o}\,2})\cdot { q}_{\rm o}
  ({ x}+{ x}_{0};N_{{\rm o}\,2}),
 \label{nonsym}
 \end{equation}
where $\delta\varphi=\varphi- {\varphi}'$,  $\delta { v} = { v} -{ v}'$, $x_{0}=v'\tau $,
and $\tau$  is the (initial) time delay.  
As the second packet moves in this frame of reference,
the 
regime of 
strong nonlinear interaction between the condensates is  (theoretically) 
unavoidable. 
Note that, even  before collision,  a
time-dependent interference term in  $\vert  \psi_{\rm o  }(x,t) \vert^{2}$ 
begins to influence the packet dynamics, see Fig. 4. 
For example,  
the r.h.s. of  (\ref{timedep1DD}) contains 
\begin{equation}
\sim \partial_{x}\left\{\, 
2\,{\rm cos}(m \,\delta { v} x - \delta\omega \,t + \delta\varphi )\, 
\phi_{\rm o}({ x}; N_{{\rm o}\,1})\,\phi_{\rm o}( x-\delta { v}t + x_{0}; N_{{\rm o}\,2})
\,\right\},
\label{fringes}
\end{equation}
where 
$$ 
\hbar \delta \omega = m\,\delta { v} (v + v')/2 - \left(\,\vert\tilde{\mu}\vert(N_{{\rm o}\,2}, v)- \vert\tilde{\mu}\vert(
N_{{\rm o}\,1}, v')\,\right)
$$
and $\vert\tilde{\mu}\vert \propto (\,N_{\rm o}/N_{\rm o}^{*})^{2}\,\vert\tilde{\varepsilon}_{0}(v)  \vert^{2}$.
The ratio $\vert\tilde{\mu}_{2}\vert /\vert\tilde{\mu}_{1}\vert$ can be of the order
of $10^{1}$, and  the characteristic scale of fringes (\ref{fringes}) is   
$$
\frac{ \pi (\hbar/m\,\delta v)}{ L_{0} } \simeq (10\sim 30)\sqrt{
\frac{ \vert\tilde{\mu}\vert(N_{{\rm o}\,1}, v)}{ (m/M)\, M c_{l}^{2}/2}
}\simeq  5 \sim 10, 
$$
that is they are of the long-wavelength nature.

To answer the question which model (conserving (\ref{timedep1D}),(\ref{timedep1DD})
or kinetic \cite{Keldysh},\cite{excitoner} one) is more adequate to describe the packet collision, we
have to compare the estimate of interaction time, 
$$
\tau ^{*} \simeq  L_{{\rm ch}\,2}/\delta v \sim 10^{3}\,\tilde{a}_{B} / 0.2\,c_{l} 
\simeq  10^{-9} \sim 10^{-10}\,{\rm s}, 
$$
and characteristic time scales of the processes
\begin{equation}
\vert N_{{\rm o}\,2}(t), v\rangle \cdot \vert N_{{\rm o}\,1}(t), v' \rangle \, \rightarrow \,\vert N_{{\rm o}\,2}(t)\pm \delta N, v \rangle \cdot
\vert N_{{\rm o}\,1}(t) \mp \delta N, v' \rangle
\label{process}
\end{equation} 
driven by phonons or by $x-x$ interaction. 
Note that some thermal  
phonons have to be excited  in the system to assist such transitions,  and   
the value of $\tau ^{*}$ is  of the order of scattering time of the  exciton-LA-phonon interaction
(although without any macroscopical occupancy) \cite{Ivanov}.

If  processes  (\ref{process}) are driven by the lattice phonons,
two phonons are necessary to satisfy the laws of conservation.
For instance,  we choose $\delta N = +1$  in  (\ref{process}) and obtain 
(see Eqs. (\ref{cond_and_exc1}),(\ref{cond-and-exc1}))  
$$
\hbar k_{1,\,x}= m \delta v +  3Mv\,\vartheta(N_{{\rm o}\,2},\,v) - 3Mv'\,\vartheta(N_{{\rm o}\,1},\,v') + \hbar  k_{2,\,x},
$$
$$
\hbar c_{l} \vert k_{1,\,x} \vert = m \delta v\,(v+v')/2   -  3\left (\,\vert\tilde{\mu}\vert(N_{{\rm o}\,2},\,v) - 
\vert\tilde{\mu}\vert(N_{{\rm o}\,1},\,v')  + \nu_{0}\,\Phi_{\rm o\,2}^{2}/3 - \nu_{0}\,\Phi_{\rm o\,1}^{2}/3
\right)
\,+
$$
$$
+\,
3/2\,M(c_{l}^{2} +v^{2})\,\vartheta(N_{{\rm o}\,2},\,v) - 3/2\,M(c_{l}^{2} + v'^{2})\,\vartheta(N_{{\rm o}\,1},\,v') 
+
\hbar c_{l}\vert  k_{2,\,x} \vert.
$$
 Although the second packet moves faster, $m \delta v > 0 $, this state can be considered as a more 
stable  (and, thus, more preferable) one  for the excitons of the slow packet. Indeed,
the following inequality  for the effective difference between the generalized chemical potentials  
seems to be valid  
\begin{equation}
m \delta v\,(v+v')/2   -  3\left (\,\vert\tilde{\mu}\vert(N_{{\rm o}\,2},\,v) - \vert\tilde{\mu}\vert(
N_{{\rm o}\,1},\,v')  + \nu_{0}\,\Phi_{\rm o\,2}^{2}/3 - \nu_{0}\,\Phi_{\rm o\,1}^{2}/3
\right) <0,
\label{Difference}
\end{equation}
and the absolute value of the l.h.s. of (\ref{Difference}) is $ \sim \vert\tilde{\mu}\vert(N_{{\rm o}\,2}\,v)$. 
\,Thus,  within the  quantum kinetic model,  
the relevant transition probabilities  
have to be calculated at least in the second order of perturbation theory,  f.ex., 
$$
 \vert N_{{\rm o}\,2}, v\rangle 
\cdot \vert N_{{\rm o}\,1}, v' \rangle \, \stackrel{{\rm ph}_{1} }{\longrightarrow} \,
\vert N_{{\rm o}\,2},  v\rangle 
\cdot \vert N_{{\rm o}\,1},\,\hbar \omega_{{\rm ex},\,k_{j}}
,\,v' \rangle \, \stackrel{{\rm ph}_{2} }{\longrightarrow} \,
\vert N_{{\rm o}\,2} + 2, v \rangle \cdot
 \vert N_{{\rm o}\,1}-2, v' \rangle.
$$
As a result, the system of two quantum 
Boltzmann equations will describe the interaction process.   

Unlike the Boltzmann equations, 
Eqs. (\ref{timedep1D},\,\ref{timedep1DD})
contain information about the quantum coherence between two condensates explicitly
and, moreover,  can describe the case of strong interaction. 
Unlike the 1D NLS equation that supports many-soliton solutions,  Eqs. (\ref{timedep1D}),(\ref{timedep1DD})
are, in fact, quasi-1D ones, and $\nu_{0}>0$ in (\ref{timedep1D}).
Therefore, it is an open question what happens  with two (excitonic) solitons 
after they collide in the crystal.  

In this work, we assume that the dominant process(es) of condensate interactions is that one(s)
leading to 
$\partial_{t} N_{{\rm o}\,2}>0$. Then, at the time scales $ \gg \tau^{*}$, 
one solitonic packet  can appear as a result of these processes.
Such a resultant ballistic packet 
can be approximately described by the steady-state  one-soliton solution of Eqs. (\ref{eq111}),(\ref{eq222})
with 
\,$\tilde{N}_{\rm o} \approx  N_{{\rm o}\,2} + N_{{\rm o}\,1}$ 
and the low of energy conservation,
\begin{equation}
E(N_{{\rm o}\,1}, v') + E(N_{{\rm o}\,2}, v) \approx E(\tilde{N}_{\rm o}=N_{{\rm o}\,2} + N_{{\rm o}\,1}, \tilde{v}).
\label{ENergy}
\end{equation}
If $\tilde{N}_{\rm o} < N_{\rm o}^{*}$, all the approximate solutions having been found in this study are valid  to
describe the resultant packet.

As we prescribed the value of $\tilde{N}_{\rm o}$, we 
have to estimate the value of  $\tilde{v}$ from Eq. (\ref{ENergy}), (generally, \,$\tilde{v} \ne  v$).
Moreover, we have to assume that the total momentum of the condensates, 
$P_{x}(N_{{\rm o}\,1}, v') + P_{x}(N_{{\rm o}\,2}, v)$,  may not be conserved because of  
lattice participation in such a condensate ``merger''. 
However, the challenging question of the  
results of coherent packet collision needs further theoretical and experimental
efforts.

\section{Conclusion}
 
In this study, we considered a model within which  
the inhomogeneous excitonic {\it condensate} with a nonzero momentum can be investigated. 
The important physics we include in our model is 
the exciton-phonon interaction
and the appearance of 
a coherent part of the crystal displacement field, which  renormalizes 
the x-x interaction vertices.
Then, the condensate wave function and its energy
can be calculated exactly in the simplest quasi-1D model,
and the solution is a sort of Davydov's  soliton \cite{Davydov}.
We believe that the transport and other unusual properties 
of the coherent para-exciton
packets in Cu$_{2}$O
can be described in the framework of the proposed model properly generalized 
to meet more realistic conditions. 

We showed that there are two critical velocities in the theory, 
namely, $v_{\rm o}$ and $v_{\rm cr}$. 
The first one, $v_{\rm o}$, comes from the renormalization of two particle exciton-exciton interaction due to phonons,
and the bright soliton state can be formed if $v>v_{\rm o}$.
Then, the important parameter, which  controls the shape and the characteristic width of 
the condensate wave function, 
is \,$\vert\tilde{\mu}\vert /\mu^{*}$, see Eqs. (\ref{MU1}),(\ref{need2}). 
The second velocity, $v_{\rm cr}$, comes from use of Landau arguments \cite{Fetter} for investigation of the dynamic
stability\,/\,instability of the moving condensate. In fact,  the important parameter, which controls
the emission of excitations, is \,$\vert\tilde{\mu}\vert /\mu_{\rm cr}$, see Eq. (\ref{cr_mu}).
Then, within 
the semiclassical approximation for the condensate excitations, we found 
more close $v$ is to $c_{l}$ 
($\mu_{\rm cr} < \vert \tilde{\mu }\vert (N_{\rm o},\,v)< \mu^{*}$) more stable the  coherent packet is.
It is interesting to discuss the possibility of observation of an instability when the condensate can be formed
in the inhomogeneous state with $v \ne 0 $, but with $ v_{\rm o} <v< v_{\rm cr}(N_{\rm o})$, or, better, 
$\vert \tilde{\mu }\vert (N_{\rm o},\,v) < \mu_{\rm cr}$.
Such a coherent packet has to disappear during
its move through a single pure crystal used for experiments. As the shape of the moving packet depends on time,
the form of the registered signal may depend on the crystal length
changing from the solitonic to the standard diffusion density profile.

We found that the excited states of the moving
exciton-phonon condensate can be described by use of the language of elementary excitations.
 Although
the possibility of their direct observation
is an unclear question itself (and, thus, the question about the gap in the excitation spectrum is still an open one),
the stability 
conditions of 
the  moving condensate 
can be derived from the low-energy asymptotics of the excitation spectra at $T\ll T_{c}$.
However, the stability problem is not without difficulties
\cite{book},\cite{Rou}.  
One can easily imagine the situation when
the condensate moves in a very high quality crystal, but with some impurity region 
prepared, f.ex., in the middle of the sample. In this case, the excitonic superfluidity can be examined  
by impurity scattering of the  ballistic condensate. 
Indeed,  such impurities
could bound the noncondensed excitons, which always accompany the condensate, and could mediate, for instance, the
emission of the outside excitations. 
The last process may lead to depletion of the condensate and, perhaps,
some other observable effects, such as damping, bound exciton PL, etc.. 


\section{Acknowledgements}

One of the authors (D.R.) thanks I.~Loutsenko for helpful
discussions, A.~Mysyrowicz for useful comments, 
and E.~Benson and E.~Fortin for  
providing the results  
of their work before publishing.


\begin{thebibliography}{100}


\bibitem{review}
J. P. Wolfe, J. L. Lin,
D. W. Snoke and by A. Mysyrowicz in {\it Bose-Einstein
Condensation}, edited by
A. Griffin, D. W. Snoke and S. Stringari (Cambridge University Press,
Cambridge, 1995).

\bibitem{Butov} 
L. V. Butov, A. Zrenner, G. Abstreiter, G.~Bohm, and
 G.~Weimann, Phys. Rev. Lett {\bf 73}, 304, (1994); Physics--Uspekhi {\bf 39},
751 (1996);
\newline
L. V. Butov, A. I. Filin, Phys. Rev. B {\bf 58}, 1980 (1998).

\bibitem{Snoke} 
V. Negoita, D. W. Snoke, and K. Eberl, cond-mat/9901088

\bibitem{Lin}
J. L. Lin, J. P. Wolfe, Phys. Rev. Lett. {\bf 71}, 122
(1993).

\bibitem{Goto}
T. Goto, M. Y. Shen, S. Koyama, T. Yokouchi, Phys. Rev. B
{\bf 55}, 7609 (1997). 
 
\bibitem{Fortin}
E. Fortin, S. Fafard, A. Mysyrowicz, Phys. Rev. Lett.
{\bf 70}, 3951 (1993).

\bibitem{Benson}
E. Benson, E. Fortin, A. Mysyrowicz, Phys. Stat. Sol. B {\bf 191}, 345 (1995);
Sol. Stat. Comm. {\bf 101}, 313, (1997). 

\bibitem{Hanamura1} 
E. Hanamura, Sol. Stat. Comm. {\bf 91}, 889 (1994);
\newline
J.~Inoue, E.~Hanamura, Sol. Stat. Comm. {\bf 99}, 547 (1996);

\bibitem{Fernandez}
J. Fern\'andez-Rossier and C. Tejedor, Phys. Rev. Lett. {\bf 78}, 4809 (1997);
\newline 
J. Fern\'andez-Rossier, C. Tejedor,  R. Merlin, cond-mat/9909232.

\bibitem{Tichodeev}
A. E. Bulatov, S. G. Tichodeev, Phys. Rev. B {\bf 46}, 15058 (1992);
\newline
G. A. Kopelevich, S. G. Tikhodeev, and N. A. Gippius, JETP {\bf 82}, 1180 (1996).

\bibitem{Loutsenko}
I. Loutsenko, D. Roubtsov, Phys. Rev. Lett. {\bf 78},
3011 (1997);
\newline
D. Roubtsov, Y. L\'epine, Phys. Stat.  Sol. (B) {\bf 210}, 127 (1998); cond-mat/9807140.  

\bibitem{Sham}
Th. \"Ostreich, K. Sch\"onhammer, L. J. Sham, 
Phys. Rev. Lett.  {\bf 74}, 4698 (1995); cond-mat/9807135.

\bibitem{boser}
A.~Imamo\=glu and R.~J.~Ram, Phys.  Lett.  A {\bf 214},
193 (1996);
\newline
W. Zhao, P. Stenius, A.~Imamo\=glu, Phys. Rev. B {\bf 56}, 5306 (1997). 

\bibitem{Kavoulakis}
G.~M.~Kavoulakis, G.~Baym, and J.~P.~Wolfe, Phys. Rev. B {\bf 
53}, 7227 (1996);
\newline
G.~M.~Kavoulakis, Y.-C. Chang, and G.~Baym, Phys. Rev. B {\bf 
55}, 7593 (1997);

\bibitem{Lozovik} 
Yu. E. Lozovik, A. V. Poushnov, JETP {\bf 88}, 747 (1999);
cond-mat/9803318.

\bibitem{discussion}
S. G. Tichodeev, Phys. Rev. Lett. {\bf 78}, 3225 (1997);
\newline A.~Mysyrowicz, {\it ibidem}, 3226 (1997).

\bibitem{Ivanov}
A. L. Ivanov, C. Ell, and H. Haug, Phys. Rev. E {\bf 55}, 6363 (1997);
Phys. Rev. B {\bf 57}, 9663 (1998).

\bibitem{Schmitt-Rink} S.~Schmitt-Rink, D.~S.~Chemla, and
D.~A.~B.~Miller, Adv. Phys. {\bf 38}, 89 (1989).

\bibitem{Keldysh}
L.~V.~Keldysh and A.~N.~Kozlov, Sov.  Phys.  JETP
{\bf 27}, 521 (1968);
\newline
E. Hanamura, H. Haug, Phys. Rep. C {\bf 33}, 209 (1977).

\bibitem{ScLength}
J. Schumway and D. M. Ceperley, cond-mat/9907309, submitted to Phys. Rev. B.

\bibitem{Popov}
V. N. Popov, {\it Functional Integrals and Collective Modes}, (Cambridge University Press, 
N.Y., 1987).

\bibitem{miscellaneous} 
M. F. Miglei, S. A. Moskalenko, and A. V. Lelyakov, Phys. Stat. Sol. {\bf 35}, 389
(1969);
\newline V. M.~Nandkumaran and K. P.~Sinha, Zeit. f\"ur Phys. B {\bf 22}, 173 (1975).

\bibitem{Davydov} 
A. S. Davydov, {\it Solitons in Molecular Systems}, 
(Reidel, 1984).

\bibitem{Griffin}
A.~Griffin, Phys. Rev. B {\bf 53}, 9341 (1996); cond-mat/9901172.

\bibitem{Hu}
G. Huang and B. Hu,  Phys. Rev. B {\bf  58},  9194 (1998).

\bibitem{Fetter} 
Al. L. Fetter and J. D. Walecka, {\it Quantum Theory of Many-Particle 
System}, (McGrav-Hill, New York, 1971).

\bibitem{Pit}
S. Giorgini, L. P. Pitaevskii, S. Stringari, Phys. Rev. A  {\bf 54}, 4633 (1996);
\newline
F. Dalfovo, S. Giorgini, L. P. Pitaevskii, and S. Stringari, 
Rev. Mod. Phys. {\bf 71}, 463 (1999); cond-mat/9806038. 

\bibitem{Hopfield} 
J.J. Hopfield, Phys. Rev. {\bf 112}, 1555 (1958).

\bibitem{Keldysh1}
A. L. Ivanov, H. Haug, and L.~V.~Keldysh, Phys. Reports {\bf 296},
237 (1998);
\newline
H. Haug, A. L. Ivanov, and L.~V.~Keldysh, {\it Nonlinear Optical
Phenomena in Semiconductors and Semiconductor Microstructures}, 
(World Scientific, 1999).

\bibitem{book}  
S. A. Moskalenko, D. W. Snoke,
{\it Bose Condensation of Excitons and Coherent Nonlinear Optics}, 
(Cambridge University Press, in press).

\bibitem{interfer} 
E.~Benson, E.~Fortin, B. Prade, and A.~Mysyrowicz, Europhys. Lett. {\bf 40}, 
311 (1997).

\bibitem{excitoner}
A.~Mysyrowicz, E.~Benson, and E.~Fortin, 
Phys. Rev. Lett.  {\bf 77}, 896 (1996).

\bibitem{Rou}
D. Roubtsov, Y. L\'epine, Phys. Lett. A {\bf 246}, 139 (1998); 
cond-mat/9807023.

\bibitem{somebook} 
H. Stolz, {\it Time-Resolved Light Scattering from Excitons}, 
(Springer-Verlag, 1994).


\end{thebibliography}
\end{document}